\documentclass[11pt]{article}
\usepackage{latexsym,amssymb,amsmath,times}
\usepackage{enumitem}
\usepackage{float}

\usepackage{hyperref}
\usepackage[square,sort&compress,comma,numbers]{natbib}
\makeatletter

\@addtoreset{equation}{section} \makeatother

\input amssym.def
\input amssym.tex

\usepackage{tikz}
\usetikzlibrary{decorations.markings}

\newcommand{\half}{{{\textstyle\frac{1}{2}}}}

\newcommand{\be}{\begin{equation}}
\newcommand{\ee}{\end{equation} }
\newcommand{\beqa}{\begin{eqnarray} }
\newcommand{\eeqa}{\end{eqnarray} }
\newcommand{\ba}{\begin{array}}
\newcommand{\ea}{\end{array}}
\newcommand{\bpm}{\begin{pmatrix}}
\newcommand{\epm}{\end{pmatrix}}

\newcommand{\Spin}{\mathbf{Spin}}

\newcommand{\rmd}{{\rm d}}

\newcommand{\rmC}{{\rm C}}

\newcommand{\mbg}{\mathbf{g}}

\newcommand\hcL{{\hat{\cal L}}}

\newcommand{\ODD}{\mathbf{O}(D,D)}

\newcommand{\Spint}{{\Spin(1,9)}}
\newcommand{\oSpint}{{{\Spin}(9,1)}}

\newcommand\Tr{{\rm Tr}}

\newcommand\cD{{\cal D}}

\newcommand\cF{{\cal F}}
\newcommand\cG{{\cal G}}
\newcommand\cH{{\cal H}}

\newcommand\cJ{{\cal J}}

\newcommand\cL{{\cal L}}
\newcommand\cM{{\cal M}}

\newcommand\cP{{\cal P}}

\newcommand\cS{{\cal S}}

\newcommand\DFT{{\scriptscriptstyle{\rm DFT}}}

\newcommand\NS{{\scriptscriptstyle{\rm NSNS}}}

\newcommand\dis{\displaystyle}

\newcommand\seceq{=}

\def\brP{\bar{P}}

\def\brcP{\bar{\cP}}

\newcommand{\DO}{\mathbf{\nabla}}

\newcommand{\na}{{\nabla}}

\newcommand\YM{\rm\scriptscriptstyle{YM}}
\newcommand\gYM{g_{\rm \scriptscriptstyle{YM}}}

\begin{document}
\begin{titlepage}
\title{
\vskip 2cm  $\ODD$ Covariant Noether Currents and Global Charges \\
in \\
Double Field Theory  \\
~}

\author{\sc Jeong-Hyuck Park${}^{\dagger}$, Soo-Jong Rey${}^{\sharp, \star}$, Woohyun Rim${}^{\sharp}$ and Yuho Sakatani${}^{\sharp}$}
\date{}
\maketitle \vspace{-1.0cm}
\begin{center}
~\\
${}^{\dagger}$Department of Physics, Sogang University,  Seoul 121-742 KOREA\\
${}^{\sharp}$School of Physics and Astronomy,
Seoul National University,  Seoul 08862 KOREA\\
${}^{\star}$Fields, Gravity \& Strings, Center for Theoretical Physics of the Universe \\
Institute for Basic Sciences, Daejeon 34047 KOREA\\
~~~\\~\\
\end{center}
\begin{abstract}
\vskip0.2cm
\noindent
Double field theory is an approach for massless modes of string theory, unifying and geometrizing all gauge invariance in manifest $\mathbf{O}(D,D)$ covariant manner. In this approach, we derive off-shell conserved Noether current and corresponding Noether potential associated with unified gauge invariance. We add Wald-type counter two-form to the Noether potential and define conserved global charges as surface integral. We check our $\mathbf{O}(D,D)$ covariant formula against various string backgrounds, both geometric and non-geometric. In all cases we examined, we find perfect agreements with previous results. Our formula facilitates to evaluate momenta along not only ordinary spacetime directions but also dual spacetime directions on equal footing. From this, we confirm recent assertion that null wave in doubled spacetime is the same as macroscopic fundamental string in ordinary spacetime.
\end{abstract}


\thispagestyle{empty}

\end{titlepage}
\newpage
\tableofcontents 
\newpage
\rightline{\sl For out of olde feldes, aas men seith, }
\rightline{\sl Cometh al this newe corn fro yeer to yere;}
\rightline{\sl And out of olde bokes, in good feith,}
\rightline{\sl Cometh al this newe science that men lere.}
\rightline{ --- Geoffrey Chaucer}
\vskip1cm

\section{Introduction and Summary}
String theory is known to possess enormous (possibly infinite-dimensional) symmetry that goes beyond the scope of conventional field theories. Double Field Theory (DFT)~\cite{Siegel:1993th,Hull:2009mi,Hull:2009zb,Hohm:2010jy,Hohm:2010pp}   is a new approach for keeping manifest the ($\mathbb{R}$-valued extension of) $\ODD$ T-duality symmetries as well as unifying and geometrizing all gauge invariance of the massless fields in string theory. It dispenses full-fledged string theory while retaining theory's most salient features, so the approach offers a novel and effective method for investigating stringy characters and gaining physical insights to intricacies of the string theory.

The novelty of DFT is doubling the spacetime dimensions at the benefit of geometrizing all gauge invariance of the graviton, Kalb-Ramond, and Yang-Mills fields (further the equations of motion are invariant under a constant shift in the dilaton field). These massless fields live in this doubled spacetime; not only vacuum configuration but also physical excitations above the vacuum extend through the doubled spacetime. Yet, the theory should be constrained such that actual mechanical degrees of freedom live on a middle-dimensional subspace of the doubled spacetime.
Therefore, the DFT poses a novel question for how Noether currents associated with asymptotic symmetries of the massless fields and global charges associated with their physical excitations are measured in the doubled spacetime in $\ODD$ covariant manner \footnote{By the $\ODD$ covariance of DFT, we just mean the counterpart to the $GL(D)$ covariance of gravity.}. This work constitutes the answer we found for this question.


To be specific, we shall proceed with the Lagrangian formulation of the `heterotic' DFT, whose field contents include the NS-NS sector coupled to Yang-Mills~\cite{Jeon:2011kp,Choi:2015bga} (\textit{c.f.~}\cite{Hohm:2011ex,Bedoya:2014pma,Lee:2015kba}).  Our main result is summarized by the $\ODD$-covariant  expression of a generic \textit{conserved global charge}, spelled in Eq.(\ref{MAINFINAL}), which we copy here:
\be
\boxed{
Q_{\rm total}[X]=\oint_{\partial\cM}\rmd^{{\scriptscriptstyle{D{-2}}}} x_{AB}~e^{-2d}\left[K^{[AB]}+2X^{[A}B^{B]}
+{1 \over \gYM^{2}} \Tr\left\{12 (P\cF\brP)^{[AB}V^{C]}X_{C}\right\}\right]\,} \, .
\label{IntroMAIN}
\ee
The first two terms in the integrand correspond to the NS-NS sector DFT version of the Noether (or Komar) potential and its counter-correction \textit{a la} Wald~\cite{Wald:1993nt,Iyer:1994ys,Iyer:1995kg}. The last term is the contribution of the Yang-Mills sector that is coupled minimally to the DFT. Using this formula, we can compute conserved global charges such as mass, doubled translational momenta, and angular momenta. We could also study the asymptotic symmetry algebra, which we relegate to a future work.

In gauge and gravity theories, symmetry currents and conserved charges play important roles in analyzing classical and quantum dynamics. In the Hamiltonian formulation, the Arnowitt-Deser-Misner (ADM) \cite{Arnowitt:1962hi} approach constructs these conserved charges, which are related to Hamiltonian surface terms that should be added to constraints for well-defined Hamiltonian generators. Using this approach, their algebras were studied in a variety of contexts \cite{Henneaux:1984xu}.

In Lagrangian formulation with a well-defined action functional, which was developed after the Hamiltonian formulation, the conserved currents are derived by the Noether theorem and the global charges are obtained as hypersurface integral of the current densities. However, proper Lagrangian formulation of symmetry currents and conserved charges are often complicated by the background vacuum configuration and also by fall-off behaviour of dynamical excitations. This is because global symmetry is defined by the asymptotic symmetry set by the orbit of gauge or diffeomorphism transformations at infinity with the common asymptotic behaviour. In other words, a specific choice of the asymptotic boundary condition puts the true local symmetries to a subset of the full gauge or diffeomorphism group. If the quotient of the full gauge or diffeomorphism transformation by this subset is nontrivial, it defines the asymptotic symmetries associated with the chosen boundary conditions. For some representative works that touch on these issues in various dimensions and that deals with physical implications, see, for example, \cite{Jang:1979zz} - \cite{Rey:1996sm}.

The Lagrangian formulation for the symmetry currents and conserved charges \cite{Barnich:2001jy}
has the advantage of manifest covariance. In this formulation, however, the action may in general contain boundary terms, which play an important role. Though not contributing to the equations of motion, the surface terms contribute to setting boundary conditions. Accordingly, the asymptotic symmetries depend sensitively on the boundary terms in the action. Here is a good place to recall them in the context of the ordinary metric formulation of gravity.
The well-defined variational principle that yields the Einstein's equation is provided by the Einstein-Hilbert action,
\be
I = {1 \over 16 \pi G_N}\,\biggl[
\int_{\Sigma_d} \sqrt{-g} R(g) + 2 \oint_{\partial \Sigma_d} \sqrt{|h|} (K - K_0)\biggr]\,.
\ee
Here, $\Sigma_d$ is a general pseudo-Riemannian manifold with metric $g$ as the second fundamental form, $\partial \Sigma_d$ is a (spacelike, lightlike or timelike) boundary of $\Sigma_d$ with induced metric $h$. The surface term, the Gibbons-Hawking term,\footnote{See \cite{Berman:2011kg} for a DFT extension of Gibbons-Hawking term.} allows its variation to counter off variations of the derivative of the metric so that Dirichlet boundary condition for the metric suffices.

The action is not just for facilitating equations of motion as the condition for stationary configuration. Associated with large diffeomorphism gauge transformations, one can construct conserved Noether currents and global charges from the action. The global gauge transformations have support at asymptotic boundary $\partial M_{d-1}$ of a timelike or lightlike hypersurface $M_{d-1}$ in $\Sigma_d$. However, in order to be able to express them as surface integrals at $\partial M_{d-1}$, the metric should additionally obey Dirichlet boundary condition at $\partial M_{d-1}$. For instance, in asymptotically flat spacetime, the action must maintain the property that a physical excitation belongs to stationary configuration under all possible variations preserving asymptotic flatness. 

With the Gibbons-Hawking surface term alone, the action renders its on-shell variation a non-zero surface term.
However, the hypersurface where the requisite counterterm is embedded is forced to fluctuate and so its variation is no longer well-defined. This difficulty can be lifted by replacing the Gibbons-Hawking term by a new counterterm which is a local function of boundary metric and Ricci curvature only and hence independent of an embedding. A concrete proposal along this direction was put forward in asymptotic flat spacetime~\cite{Mann:2005yr} - \cite{Mann:2008ay}
and in asymptotic (anti) de Sitter spacetime~\cite{Fischetti:2012rd}.
Such counterterm renders a well-posed variational principle for all deformations of the metric consistent with asymptotic flatness. Furthermore, such a counterterm also facilitates the computation of conserved global charges in a procedure similar to that of Brown and York~\cite{Brown:1992br}.

To explain the Noether currents and conserved global charges,  \textit{i.e.} (\ref{IntroMAIN}), in a self-contained manner and to apply them to various known string theory backgrounds, we organized this paper as follows. In section~\ref{SECreview}, we first review the semi-covariant formulation of DFT, closely following the formulation developed in \cite{Jeon:2010rw,Jeon:2011cn,Park:2013mpa,Lee:2013hma}. The formalism based on the semi-covariant derivative and its complete covariantization is essential to decode our  results. Section~\ref{SECMAIN} contains main results of this paper. We first derive an off-shell conserved Noether current, $J^{A}$~(\ref{bareJ}), and corresponding  Noether potential, $K^{AB}$~(\ref{Kpotential}), which originates from the generalized diffeomorphism invariance of DFT.  Meanwhile, we also identify  the DFT extension  of the off-shell conserved \textit{Einstein curvature tensor}, $\na_{A}G^{AB}=0$~(\ref{deftwoindexJ}). We then follow  the prescription by Wald~\cite{Wald:1993nt,Iyer:1994ys,Iyer:1995kg}~(see also
\cite{Kim:2013zha,Hyun:2014kfa}).   Introducing  counter corrections,  we  modify the off-shell conserved Noether current and the potential,   $\widehat{J}^{A}$~(\ref{Jpr}), $\widehat{K}^{AB}$~(\ref{Kpr}).  The integration of the  modified Noether current  defines the  $\ODD$-covariant conserved global charge. After finishing our analysis on  the pure NS-NS sector DFT, we generalize to include
Yang-Mills~\cite{Jeon:2011kp,Choi:2015bga} as well as  cosmological constant. In section~\ref{SECAPP}, we apply our general result to various known backgrounds, which include  fundamental string~\cite{Dabholkar:1990yf},  Reissner-Nordstr\"{o}m black hole, black five-brane~\cite{Horowitz:1991cd} and  linear dilaton background~\cite{Maldacena:1997cg}. We find  perfect agreement. We further consider genuine DFT (or stringy)  backgrounds such as null-wave in doubled spacetime~\cite{Berkeley:2014nza} and non-Riemannian background~\cite{Lee:2013hma}. We evaluate their momenta along not only ordinary spatial or temporal directions but also dual directions. We  confirm the assertion of Berkeley, Berman and Rudolph~\cite{Berkeley:2014nza} that a massless null wave in doubled spacetime is identifiable with a macroscopic fundamental string in ordinary spacetime.

In the Appendix, we relegated some useful technical formulae and detailed analysis of the asymptotic fall-off behaviour at infinity.\\

\textit{Note added}:  While we were finishing this paper, we became aware of the work by C. Blair~\cite{BlairOverlap},
which also studies conserved charges in DFT.\\

\section{Review: Semi-Covariant
Formulation of Double Field Theory\label{SECreview}}

We begin with self-contained review of the semi-covariant formulation of  the bosonic DFT for the NS-NS sector ~\cite{Jeon:2010rw,Jeon:2011cn} and also the Yang-Mills sector~\cite{Jeon:2011kp,Choi:2015bga}. They constitute the massless modes of string theory at leading order in string coupling perturbation theory. For further extensions beyond the leading order, we refer readers to  \cite{Jeon:2011vx} for fermions, \cite{Jeon:2012kd} for the R-R sector, and \cite{Jeon:2012hp,Jeon:2011sq,Cho:2015lha} for the (gauged) maximal and half-maximal supersymmetric completions.\footnote{In particular,  thanks to the twofold spin groups, \textit{i.e.~}$\Spint\times\oSpint$, the  distinction between  IIA and IIB disappears~\cite{Jeon:2012hp}, and the maximal $D=10$  supersymmetric double field theory   \textit{unifies} IIA and IIB supergravities. }

The DFT is defined over the doubled, $(D+D)$-dimensional spacetime. Denote the $\ODD$ vector indices by capital Latin letters, $A,B,C,\cdots=1,2,\cdots,D{+D}$. There exists a unique $\ODD$ invariant constant metric,
\begin{equation}
\cJ_{AB}=\left[ \begin{matrix} \ 0\ &\ 1\ \\ \ 1\ & \ 0 \ \end{matrix} \right]\,.
\label{ODDeta}
\end{equation}
Using this invariant metric, we freely raise and lower the $\ODD$ tensor indices.

The actual physics is realized in $D$-dimensional subspace.
As the DFT starts with doubled $(D+D)$-dimensional spacetime, this doubled spacetime must be projected appropriately. We do this by imposing the property that the doubled coordinate system satisfies {\sl local equivalence relations}~\cite{Park:2013mpa,Lee:2013hma},
\be
x^{A}~\simeq ~x^{A}+\phi(x) \partial^{A}\varphi (x) \,,
\label{aCGS}
\ee
which was termed as `coordinate gauge symmetry'. In (\ref{aCGS}),  $\phi(x)$ and $\varphi(x)$ are arbitrary smooth functions in DFT. Each equivalence class or each gauge orbit defined by the equivalence  relation~(\ref{aCGS}) represents a single physical point, and diffeomorphism invariance refers to a symmetry under arbitrary reparametrizations of the gauge orbits.

The equivalence relation (\ref{aCGS}) is realized in DFT  by enforcing that arbitrary  functions and their arbitrary  derivatives, denoted here collectively  by $\Phi$, are  invariant under the coordinate gauge transformations \textit{shift},
\be
\ba{ll}
\Phi(x+\Delta)=\Phi(x)\,,\quad&\quad\Delta^{A}=\phi\partial^{A}\varphi\,.
\ea
\label{aTensorCGS}
\ee
The coordinate gauge symmetry can be also realized as a local Noether symmetry on a string worldsheet~\cite{Lee:2013hma}.

The symmetry under the coordinate gauge transformation~(\ref{aTensorCGS}) is equivalent  (\textit{i.e.~}necessary~\cite{Park:2013mpa} and sufficient~\cite{Lee:2013hma}) to the {\sl section condition}~\cite{Hohm:2010pp},
\be
\partial_{A}\partial^{A}= 0\,.
\label{aseccon}
\ee
Acting on arbitrary functions, $\Phi$, $\Phi^{\prime}$,  and their products, the section condition leads to
\be
\ba{ll}
\partial_{A}\partial^{A}\Phi\seceq 0\quad(\mbox{weak~~constraint})\,,\quad&\quad
\partial_{A}\Phi\partial^{A}\Phi^{\prime}\seceq 0\quad(\mbox{strong~~constraint})\,.
\ea
\label{aseccon2}
\ee

Diffeomorphism transformation in the doubled-yet-gauged coordinate system is generated by a generalized Lie derivative~\cite{Siegel:1993th,Gualtieri:2003dx,Grana:2008yw}. Acting on $n$-indexed field, it is defined by
\be
\hcL_{X}T_{A_{1}\cdots A_{n}}:=X^{B}\partial_{B}T_{A_{1}\cdots A_{n}}+\omega_{{\scriptscriptstyle{T\,}}}\partial_{B}X^{B}T_{A_{1}\cdots A_{n}}+\sum_{i=1}^{n}(\partial_{A_{i}}X_{B}-\partial_{B}X_{A_{i}})T_{A_{1}\cdots A_{i-1}}{}^{B}{}_{A_{i+1}\cdots  A_{n}}\, .
\label{tcL}
\ee
Here, $\omega_{{\scriptscriptstyle{T\,}}}$ denotes the weight of the $T$ field.
In particular, the generalized Lie derivative of the $\ODD$ invariant metric is trivial,
\be
\hcL_{X}\cJ_{AB}=0\,.
\label{acompcJ}
\ee
The commutator of the generalized Lie derivatives is closed by the C-bracket~\cite{Siegel:1993th,Hull:2009zb},
\be
\ba{ll}
\left[\hcL_{X},\hcL_{Y}\right]=\hcL_{[X,Y]_{\rmC}}\,,\quad&\quad
[X,Y]^{A}_{\rmC}= X^{B}\partial_{B}Y^{A}-Y^{B}\partial_{B}X^{A}+\half Y^{B}\partial^{A}X_{B}-\half X^{B}\partial^{A}Y_{B}\,.
\ea
\label{agB2}
\ee

In the NS-NS sector, dynamical contents of the DFT consist of the dilaton, $d(x)$, and a pair of the projection fields $P_{AB}, \brP_{AB}$, obeying the properties
\be
\ba{llll}
P_{AB}=P_{BA}\,,\quad&\quad \brP_{AB}=\brP_{BA}\,,
\quad&\quad P_{A}{}^{B}P_{B}{}^{C}=P_{A}^{~C}\,,
\quad&\quad \brP_{A}{}^{B}\brP_{B}{}^{C}=\brP_{A}^{~C}\,.
\ea
\ee
Further, the projection fields are orthogonal and complementary:
\be
\ba{ll}
P_{A}{}^{B}\brP_{B}{}^{C}=0\,,\quad
\brP_{A}{}^B P_{B}{}^C = 0\, , \quad P_{AB}+\brP_{AB}=\cJ_{AB}\,.
\ea
\label{aorthogonalP}
\ee
The two projection fields are not independent, since
\be
P_{A}{}^B + \brP_{A}{}^B = \cJ_{A}{}^B.
\ee
The dynamical contents are contained (in addition to the dilaton) in the difference of the projection fields
\be
P_{AB}-\brP_{AB}=\cH_{AB}.
\ee
This is the well-known {\sl generalized metric}~\cite{Hohm:2010pp}, which can be also independently defined as a symmetric $\ODD$ element having the properties
\be
\cH_{AB}=\cH_{BA} \quad \mbox{and} \quad
\cH_{A}{}^{B}\cH_{B}{}^{C}=\delta_{A}^{~C}.
\ee
%

The projection fields and dilaton are naturally in the string frame. To facilitate the $\ODD$ invariant integral calculus, we assign the scaling weight of these fields as
\be
\ba{ll}
\omega[P] =\omega[\brP] = 0\,, \quad &\quad\omega[e^{-2d}] = 1 \,.
\ea
\label{ameasure}
\ee

The central construction of the DFT starts with the
semi-covariant derivative, 
defined by~\cite{Jeon:2010rw,Jeon:2011cn}
\be
\na_{C}T_{A_{1}A_{2}\cdots A_{n}}
:=\partial_{C}T_{A_{1}A_{2}\cdots A_{n}}-\omega_{{\scriptscriptstyle{T\,}}}\Gamma^{B}{}_{BC}T_{A_{1}A_{2}\cdots A_{n}}+
\sum_{i=1}^{n}\,\Gamma_{CA_{i}}{}^{B}T_{A_{1}\cdots A_{i-1}BA_{i+1}\cdots A_{n}}\,,
\label{asemicov}
\ee
The connection is defined by~\cite{Jeon:2011cn}:\footnote{In this review of the \textit{bosonic} DFT, we focus on the above `torsionless' connection~(\ref{Gammao}).
 Yet, in  \textit{supersymmetric} DFT, it is necessary to include torsions in order to ensure the `1.5 formalism'~\cite{Jeon:2012kd,Jeon:2011sq,Jeon:2012hp}.}
\be
\ba{ll}
\Gamma_{CAB}=&2\left(P\partial_{C}P\brP\right)_{[AB]}
+2\left({{\brP}_{[A}{}^{D}{\brP}_{B]}{}^{E}}-{P_{[A}{}^{D}P_{B]}{}^{E}}\right)\partial_{D}P_{EC}\\
{}&-\textstyle{\frac{4}{D-1}}\left(\brP_{C[A}\brP_{B]}{}^{D}+P_{C[A}P_{B]}{}^{D}\right)\!\left(\partial_{D}d+(P\partial^{E}P\brP)_{[ED]}\right)\,,
\ea
\label{Gammao}
\ee
Below, we shall elaborate the uniqueness of this connection. The semi-covariant derivative obeys the Leibniz rule and annihilates the $\ODD$ invariant constant metric:
\be
\na_{A}\cJ_{BC}=0\,.
\label{nacJ}
\ee

Unlike the Levi-Civita connection in the Riemannian Einstein gravity, the diffeomorphism transformation~(\ref{tcL}) cannot put the connection~(\ref{Gammao}) to vanish pointwise in doubled spacetime. One may view this as failure of the equivalence principle. This is not surprising since it is know that in string theory the equivalence principle no longer holds due to the Kalb-Ramond field and the dilaton field.

The semi-covariant Riemann curvature calculates the field strength of the connection (\ref{Gammao}):
\be
S_{ABCD}:=\half\left(R_{ABCD}+R_{CDAB}-\Gamma^{E}{}_{AB}\Gamma_{ECD}\right)\,.
\label{asemicovS}
\ee
Here, $R_{ABCD}$ denotes the ordinary  Riemann curvature associated with the connection:
\be
R_{CDAB}=\partial_{A}\Gamma_{BCD}-\partial_{B}\Gamma_{ACD}+\Gamma_{AC}{}^{E}\Gamma_{BED}-\Gamma_{BC}{}^{E}\Gamma_{AED}\,.
\ee

A crucial defining property of the semi-covariant Riemann curvature is that, under arbitrary variation of the connection~(\ref{Gammao}),  its variation takes the form  \textit{total derivative}~\cite{Jeon:2011cn},
\be
\delta S_{ABCD}=\na_{[A}\delta\Gamma_{B]CD}+\na_{[C}\delta\Gamma_{D]AB}\,.
\label{adeltaS4}
\ee
Further, the semi-covariant Riemann curvature satisfies precisely the same symmetric properties as the ordinary  Riemann curvature, including the Bianchi identity,
\be
\ba{ll}
S_{ABCD}=S_{[AB][CD]}=S_{CDAB}\,,\quad&\quad S_{[ABC]D}=0\,.
\ea
\ee
In addition, when projected by $P, \brP$, the semi-covariant Riemann curvature obeys identities  ~\cite{Jeon:2011cn},
\be
\ba{c}
P_{I}{}^{A}P_{J}{}^{B}\brP_{K}{}^{C}\brP_{L}{}^{D}S_{ABCD}\seceq0\,,
\quad\quad P_{I}{}^{A}\brP_{J}{}^{B}P_{K}{}^{C}\brP_{L}{}^{D}S_{ABCD}=0\,,\\
(P^{AB}P^{CD}+\brP^{AB}\brP^{CD})S_{ACBD}\seceq 0\,,\\
P_{I}{}^{A}\brP_{J}{}^{C}P^{BD}S_{ABCD}=P_{I}{}^{A}\brP_{J}{}^{C}\brP^{BD}S_{ABCD}\seceq \half P_{I}{}^{A}\brP_{J}{}^{C}S_{AC}\,.
\ea
\label{identitiesS}
\ee

As in the Riemannian case, we also define the semi-covariant Ricci curvature as the trace part of the semi-covariant Riemann curvature,
\be
S_{AC}:=S_{ABCD} \cJ^{BD} = S_{ABC}{}^{B},
\ee
However, unlike the Riemannian case, the above Bianchi identities imply that the Ricci curvature is traceless
\be
S_{A}{}^{A}=S^{AB}{}_{AB}\seceq0\,.
\label{trivialS}
\ee

The alluded connection~(\ref{Gammao}) turns out to be  \textit{the unique solution} to the following five requirements ~\cite{Jeon:2011cn}:
\begin{eqnarray}
&\na_{A}P_{BC}=0\,,\quad\quad\na_{A}\brP_{BC}=0\,,\label{acompP}\\
&\na_{A}d=-\half e^{2d}\na_{A}(e^{-2d})=\partial_{A}d+\half\Gamma^{B}{}_{BA}=0\,,\label{acompd}\\
&\Gamma_{ABC}+\Gamma_{ACB}=0\,,\label{aGBC}\\
&\Gamma_{ABC}+\Gamma_{BCA}+\Gamma_{CAB}=0\,,\label{aGABC}\\
&\cP_{ABC}{}^{DEF}\Gamma_{DEF}=0\,,\quad\quad\bar{\cP}_{ABC}{}^{DEF}\Gamma_{DEF}=0\,.\label{akernel}
\end{eqnarray}
The first two relations, (\ref{acompP}), (\ref{acompd}), are the  compatibility conditions with all the geometric objects --or the  NS-NS sector-- in DFT.
The third constraint~(\ref{aGBC}) is the  compatibility condition with the $\ODD$ invariant constant metric,  (\ref{nacJ}), which is also consistent with  (\ref{aorthogonalP}) and (\ref{acompP}). The next cyclic   property, (\ref{aGABC}),  makes  the semi-covariant derivative   compatible  with  the generalized Lie derivative as well as with  the C-bracket,
\be
\ba{ll}
\hcL_{X}(\partial)=\hcL_{X}(\na)\,,\quad&\quad
[X,Y]_{\rmC}(\partial)=[X,Y]_{\rmC}(\na)\,,
\ea
\ee
The last conditions~(\ref{akernel}) assert that the connection belongs to the kernel of the triple-projection fields $\cP_{ABC}{}^{DEF}, \bar{\cP}_{ABC}{}^{DEF}$. They are properties of the connection~(\ref{Gammao}) which completes to ensure the uniqueness.

The triple-projection fields carrying six indices,$\cP_{ABC}{}^{DEF}, \bar{\cP}_{ABC}{}^{DEF}$, used in (\ref{akernel}), are explicitly given by
\be
\ba{l}
\cP_{CAB}{}^{DEF}:=P_{C}{}^{D}P_{[A}{}^{[E}P_{B]}{}^{F]}+\textstyle{\frac{2}{D-1}}P_{C[A}P_{B]}{}^{[E}P^{F]D}\,,\\
\bar{\cP}_{CAB}{}^{DEF}:=\brP_{C}{}^{D}\brP_{[A}{}^{[E}\brP_{B]}{}^{F]}+\textstyle{\frac{2}{D-1}}\brP_{C[A}\brP_{B]}{}^{[E}\brP^{F]D}\,,
\ea
\label{P6}
\ee
which  satisfy the `projection' properties,
\be
\ba{ll}
\cP_{ABC}{}^{DEF}\cP_{DEF}{}^{GHI}=\cP_{ABC}{}^{GHI}\,,\quad&\quad
\brcP_{ABC}{}^{DEF}\brcP_{DEF}{}^{GHI}=\brcP_{ABC}{}^{GHI}\,.
\ea
\ee
They are symmetric and traceless,
\be
\ba{lll}
\cP_{ABCDEF}=\cP_{DEFABC}\,,\quad&\quad\cP_{ABCDEF}=\cP_{A[BC]D[EF]}\,,
\quad&\quad P^{AB}\cP_{ABCDEF}=0\,,\\
\brcP_{ABCDEF}=\brcP_{DEFABC}\,,\quad&\quad\brcP_{ABCDEF}=\brcP_{A[BC]D[EF]}\,,
\quad&\quad \brP^{AB}\brcP_{ABCDEF}=0\,.
\ea
\label{symP6}
\ee

The triple-projection fields describe anomalous part of   the semi-covariant derivative and  the semi-covariant  Riemann curvature under the generalized diffeomorphism transformations. From
\be
(\delta_{X}{-\hcL_{X}})\Gamma_{CAB}=2\left[(\cP+\brcP)_{CAB}{}^{FDE}-\delta_{C}^{~F}\delta_{A}^{~D}\delta_{B}^{~E}
\right]\partial_{F}\partial_{[D}X_{E]}\,,
\label{diffGamma}
\ee
it is straightforward to see that the generalized diffeomorphism  anomalies are all given by the triple-projection fields,
\be
\ba{l}
(\delta_{X}{-\hcL_{X}})\na_{C}T_{A_{1}\cdots A_{n}}\seceq
\dis{\sum_{i=1}^{n}2(\cP{+\brcP})_{CA_{i}}{}^{BDEF}
\partial_{D}\partial_{E}X_{F}\,T_{A_{1}\cdots A_{i-1} BA_{i+1}\cdots A_{n}}\,,}\\
(\delta_{X}-\hcL_{X})S_{ABCD}\seceq 2\na_{[A}\Big((\cP{+\brcP})_{B][CD]}{}^{EFG}\partial_{E}\partial_{F}X_{G}\Big)+2\na_{[C}\Big((\cP{+\brcP})_{D][AB]}{}^{EFG}\partial_{E}\partial_{F}X_{G}\Big).
\ea
\label{diffAnomaly}
\ee
Hence, one can easily project out the anomalies  through  appropriate contractions with the triple-projection fields.

The fully covariant derivatives are obtainable by further projections,
\be
\ba{ll}
P_{C}{}^{D}{\brP}_{A_{1}}{}^{B_{1}}\cdots{\brP}_{A_{n}}{}^{B_{n}}
\DO_{D}T_{B_{1}\cdots B_{n}}\,,~&~
{\brP}_{C}{}^{D}P_{A_{1}}{}^{B_{1}}\cdots P_{A_{n}}{}^{B_{n}}
\DO_{D}T_{B_{1}\cdots B_{n}}\,,\\
P^{AB}{\brP}_{C_{1}}{}^{D_{1}}\cdots{\brP}_{C_{n}}{}^{D_{n}}\DO_{A}T_{BD_{1}\cdots D_{n}}\,,~&~
\brP^{AB}{P}_{C_{1}}{}^{D_{1}}\cdots{P}_{C_{n}}{}^{D_{n}}\DO_{A}T_{BD_{1}\cdots D_{n}}\quad~~(\mbox{divergences})\,,\\
P^{AB}{\brP}_{C_{1}}{}^{D_{1}}\cdots{\brP}_{C_{n}}{}^{D_{n}}
\DO_{A}\DO_{B}T_{D_{1}\cdots D_{n}}\,,~&~
{\brP}^{AB}P_{C_{1}}{}^{D_{1}}\cdots P_{C_{n}}{}^{D_{n}}
\DO_{A}\DO_{B}T_{D_{1}\cdots D_{n}}\quad(\mbox{Laplacians})\,,
\ea
\label{covT}
\ee
Correspondingly, fully covariant Ricci curvature and fully covariant curvature scalar are~\footnote{For the torsionless connection~(\ref{Gammao}), the following identities hold as for the completely covariant scalar curvature,
\[
P^{AB}P^{CD}S_{ACBD}\seceq P^{AB}S_{ACB}{}^{C}\seceq-\brP^{AB}\brP^{CD}S_{ACBD}\seceq-\brP^{AB}S_{ACB}{}^{C}\,.
\]
However, it is the expression in (\ref{aRicciScalar}) that ensures the `1.5 formalism' in the supersymmetric DFTs with torsions~\cite{Jeon:2011sq,Jeon:2012hp}.   }
\be
\ba{ll}
\cS_{AB} := P_{A}{}^{C}\brP_{B}{}^{D}S_{CED}{}^{E}=P_{A}{}^{C}\brP_{B}{}^{D}S_{CD}=(PS\brP)_{AB}~~&\quad(\mbox{Ricci~curvature})\,,\\
\cS:=(P^{AC}P^{BD}-\brP^{AC}\brP^{BD})S_{ABCD}~~&\quad(\mbox{scalar~curvature})\,.
\ea
\label{aRicciScalar}
\ee
Because of the triviality of the semi-covariant scalar curvature, $S_{A}{}^{A}=S_{AB}{}^{AB}=0$~(\ref{trivialS}), hereafter we use the calligraphic font to denote the nontrivial Ricci curvature and scalar curvature,  `$\cS_{AB}, \ \cS\,$'.\\

A remark is in order at this point. As an alternative to the semi-covariant approach described above, from the outset, one may wish to postulate a ``perfectly well-behaving" connection, say $\widetilde{\Gamma}_{CAB}$,  such that it would transform as $(\delta_{X}{-\hcL_{X}})\widetilde{\Gamma}_{CAB}=-2\partial_{C}\partial_{[A}X_{B]}$  
instead of \eqref{diffGamma} and hence there would be no diffeomorphism anomalies like \eqref{diffAnomaly}. Yet, this is most likely not true in generic DFT, since such a ``perfect" connection cannot always be constructed solely out of the NS-NS sector.  It requires extra unphysical degrees of freedom or ``undetermined" part~\cite{Hohm:2011si}.  After projecting them out, the final results would be reduced to the semi-covariant formalism.\\

We now extend the semi-covariant DFT formulation to YM sector.
For a given Lie algebra-valued vector potential,  $V_{A}$, the semi-covariant YM field strength is defined by~\cite{Jeon:2011kp,Choi:2015bga}
\be
\cF_{AB}:=\na_{A}V_{B}-\na_{B}V_{A}-i[V_{A},V_{B}]\,.
\ee
Unlike the Riemannian torsionless case, the connections above are not cancelled but yields non-derivative and non-commutator contribution.

As seen from the generic formula~(\ref{diffAnomaly}), the semi-covariant YM field strength is not completely covariant but rather semi-covariant under diffeomorphisms. Again, the anomalous part is parametrized by the triple-projection fields,
\be
(\delta_{X}-\hcL_{X})\cF_{AB}=-2(\cP+\brcP)^{C}{}_{AB}{}^{DEF}\partial_{D}\partial_{[E}X_{F]}V_{C}\,.
\label{YMdiffanomaly}
\ee
Thus, following the general prescription~(\ref{covT}),  the completely covariant YM field strength is
\be
(P\cF\brP)_{AB}=-(\brP\cF P)_{BA}=P_{A}{}^{C}\brP_{B}{}^{D}\cF_{CD}\,.
\ee
The YM gauge transformation is realized by the action
\be
\ba{lll}
V_{A}&~\longrightarrow~& \mbg V_{A}\mbg^{-1}-i(\partial_{A}\mbg)\mbg^{-1}\,,\\
\cF_{AB}&~\longrightarrow~&\mbg\cF_{AB} \mbg^{-1}+i\Gamma^{C}{}_{AB}(\partial_{C}\mbg)\mbg^{-1}\,,\\
(P\cF\brP)_{AB}&~\longrightarrow~&\mbg(P\cF\brP)_{AB}\mbg^{-1}\,.
\ea
\label{YMgauge}
\ee
One finds that a two-derivative scalar fully invariant with respect to both the diffeomorphism and the YM gauge transformations is
\be
\Tr\left[(P\cF\brP)_{AB}(P\cF\brP)^{AB}\right]=\Tr\left[P^{AC}\brP^{BD}\cF_{AB}\cF_{CD}\right]\,.
\ee
Clearly, there appear doubled off-shell degrees of freedom  in the $(D{+D})$-component gauge potential. In order to halve them,  if wanted,
we may  impose the  `gauged  section   condition'~\cite{Choi:2015bga}:
\be
(\partial_{A}-iV_{A})(\partial^{A}-iV^{A})=0\,,
\label{secconV}
\ee
which,  along with  the original section condition~(\ref{aseccon}), implies $V_{A}\partial^{A}{=0}$, $\partial_{A}V^{A}{=0}$, $V_{A}V^{A}{=0}$. Here, it is implicit that the connections are in an irreducible representation of the fields that the covariant derivative acts on.

For consistency,  the  condition~\eqref{secconV} is  preserved  under all the symmetry transformations: $\ODD$ rotations, diffeomorphisms~(\ref{tcL}) and the Yang-Mills gauge symmetry~(\ref{YMgauge}).

We can construct the DFT action $I_\DFT$ for the NS-NS sector coupled to YM sector and cosmological constant as
\be
\ba{ll}
I_\DFT = \dis{\int_{\Sigma_{D}}}\cL_{\DFT}\,,\quad&\quad\cL_{\DFT}=\cL_{\NS}+\cL_{\YM}-2\Lambda e^{-2d}\,,
\ea
\label{cL}
\ee
where the integral is taken over a $D$-dimensional section {or their `manifold-like' patch}, $\Sigma_{D}$. Here, $\Lambda$ denotes the DFT-cosmological constant~\cite{Jeon:2011cn}. The fully invariant Lagrangian densities are given for each sector by
\be
\ba{l}
\cL_{\NS}=e^{-2d}(P^{AC}P^{BD}-\brP^{AC}\brP^{BD})S_{ABCD}=e^{-2d}\cS\,,\\
\cL_{\YM}=\gYM^{-2}e^{-2d}\Tr\left[P^{AC}\brP^{BD}\cF_{AB}\cF_{CD}\right]\,.
\ea
\label{NSYM}
\ee

At the outset, one needs to impose appropriate section condition. Up to $\ODD$ duality rotations, the solution to the section condition is locally unique. It is  a $D$-dimensional  \textit{section}, $\Sigma_{D}$,  characterized  by the independence of the dual ``winding'' coordinates,
\be
\dis{\frac{\partial~~}{\partial\tilde{x}_{\mu}}}\equiv 0\,.
\label{asection}
\ee
Here, the Greek letters are $D$-dimensional indices on the section $\Sigma_D$. In this foliation,
 the whole doubled coordinates are given by
\be
x^{A}=(\tilde{x}_{\mu},x^{\nu})\,.
\ee

To perform the `Riemannian' reduction  onto the $D$-dimensional section, $\Sigma_{D}$,  one only needs to parametrize the projection fields and the dilaton in terms of a $D$-dimensional Riemannian metric, $G_{\mu\nu}$,  an ordinary dilaton, $\Phi$, and a Kalb-Ramond two-form potential, $B_{\mu\nu}$~\cite{Hohm:2010pp},
\be
\ba{ll}
\cH_{AB}:= P_{AB}-\brP_{AB} =
 \left( \begin{array}{cc}
G^{-1}&-G^{-1}B\\
BG^{-1}~& ~G-BG^{-1}B
                 \end{array} \right) \,,\quad&\quad
               e^{-2d}=\sqrt{|G|}e^{-2\Phi}\,.
\ea
\label{aPARAN}
\ee
The DFT scalar curvature~(\ref{aRicciScalar}) reduces upon the section condition to
\be
\left.\cS\right|_{\Sigma_{D}}= R_{\scriptscriptstyle{G}}+4\Delta\Phi
-4\partial_{\mu}\Phi\partial^{\mu}\Phi-\textstyle{\frac{1}{12}}H_{\lambda\mu\nu}H^{\lambda\mu\nu}\,,
\label{aREDUCEDN}
\ee
where $H_{\lambda\mu\nu}:=3\partial_{[\lambda}B_{\mu\nu]}$.

Up to field redefinitions, Eq.(\ref{aPARAN}) is the most general parametrization of the ``generalized metric", $\cH_{AB}=P_{AB}-\brP_{AB}$, of which the upper left $D\times D$ block  is non-degenerate.

The parametrization of the doubled YM vector potential reads from \cite{Jeon:2011kp},
\be
V_{A}=\left(\ba{c}\varphi^{\lambda}\\A_{\mu}+B_{\mu\nu}\varphi^{\nu}\ea\right)\,,
\label{YMpara}
\ee
of which the $D$-dimensional vector,  $\varphi^{\lambda}$, which is in the YM adjoint representation can be put trivial upon the `gauged section condition'~(\ref{secconV})~\cite{Choi:2015bga}.  For the consequent expression of the completely covariant YM field strength in terms of $\varphi^{\lambda}$ and $A_{\mu}$, we refer readers to (3.19) and (3.21) of \cite{Jeon:2011kp}.

When the upper left $(D\times D)$ block of the generalized metric  is degenerate --where $G^{-1}$ is positioned in (\ref{aPARAN})--  the Riemannian metric ceases to exist  upon  the section, $\Sigma_{D}$~(\ref{asection}). Nevertheless,  the  $\ODD$ DFT  and a doubled sigma model~\cite{Lee:2013hma} have  no problem with describing   such a non-Riemannian background, as long as  the generalized metric    is a symmetric $\ODD$ element, satisfying  $\cH_{AB}=\cH_{BA}$ and $\cH_{A}{}^{B}\cH_{B}{}^{C}=\delta_{A}^{~C}$. We refer  readers to \cite{Lee:2013hma} for a concrete example  (see also a math literature~\cite{Garcia-Fernandez:2013gja}).



\section{Off-shell Noether Current, Noether Potential and  Global Charge\label{SECMAIN}}
In this section, we take the DFT of NS-NS sector coupled to YM sector and systematically derive off-shell conserved Noether current as well as the corresponding  Noether potential which originate from the generalized diffeomorphism gauge invariance. We then construct global charges, expressed in terms of surface integral of modified Noether current that follows from appropriate surface term in the DFT action. We shall first present the construction for the NS-NS sector and later incorporate the YM sector.

\subsection{Noether analysis on DFT of NS-NS sector\label{SECNSNS}}
Recall that dynamical field contents of the NS-NS sector DFT are the projection field $P_{AB}$ and the dilaton field $d$. Under variation of these fields, variation of the DFT Lagrangian~(\ref{NSYM}) takes the structure \footnote{Note that $\delta P_{AB}=(P\delta P\brP)_{AB}+(\brP \delta P P)_{AB}$.}
\be
\delta\cL_{\NS}=-2\delta d\,e^{-2d}\cS+4e^{-2d}(P\delta P\brP)^{AB}(PS\brP)_{AB}
+\partial_{A}\left[e^{-2d}\Theta^{A}\right]\,,
\label{varL}
\ee
where $\Theta^A$ in the last surface term denotes
\be
\Theta^{A}:=2(P^{AC}P^{BD}-\brP^{AC}\brP^{BD})\delta\Gamma_{BCD}\,.
\label{theta}
\ee

For infinitesimal variations, (\ref{varL}) yields the equations of motion for the projection field, respectively, the  dilaton field: \footnote{Note the equivalence, $ P_{A}{}^{C}\brP_{B}{}^{D}S_{CD}\simeq 0~\Leftrightarrow~ P_{(A}{}^{C}\brP_{B)}{}^{D}S_{CD}\simeq 0$.}
\be
\ba{ll}
\cS=(P^{AC}P^{BD}-\brP^{AC}\brP^{BD})S_{ABCD}\simeq 0\,,\quad&\quad  (P S \brP)_{(AB)} =  P_{(A}{}^{C}\brP_{B)}{}^{D}S_{CD}\simeq 0\,.
\ea
\label{EOM}
\ee
So, on the shell, the Lagrangian vanishes: the on-shell action would be given entirely by a surface term one may add to it.

The variation of the connection in (\ref{theta}) is given by~\cite{Jeon:2011cn},
\be
\ba{ll}
{\delta\Gamma}_{CAB}=&2P_{[A}^{~D}\brP_{B]}^{~E}\na_{C}\delta P_{DE}+2(\brP_{[A}^{~D}\brP_{B]}^{~E}-P_{[A}^{~D}P_{B]}^{~E})\na_{D}\delta P_{EC}\\
{}&-\textstyle{\frac{4}{D-1}}(\brP_{C[A}\brP_{B]}^{~D}+P_{C[A}P_{B]}^{~D})(\partial_{D}\delta d+P_{E[G}\na^{G}\delta P^{E}_{~D]})\\
{}&-\Gamma_{FDE\,}\delta(\cP+\brcP)_{CAB}{}^{FDE}\,.
\ea
\label{deltaGamma}
\ee
The last line does not contribute to (\ref{theta}) as, from (\ref{akernel}) and (\ref{symP6}), the following projection properties follow:
\be
\ba{l}
P^{CB}\Gamma_{FDE\,}\delta(\cP+\brcP)_{CAB}{}^{FDE}=-P^{CB}
(\cP+\brcP)_{CAB}{}^{FDE}\delta\Gamma_{FDE}=0\,,\\
\brP^{CB}\Gamma_{FDE\,}\delta(\cP+\brcP)_{CAB}{}^{FDE}=-\brP^{CB}
(\cP+\brcP)_{CAB}{}^{FDE}\delta\Gamma_{FDE}=0\, .
\ea
\ee
Using also the relations
\be
\ba{c}
P^{BD}\delta\Gamma_{BCD}=2P_{C}{}^{B}\partial_{B}\delta d-\na^{B}\delta P_{BC}\,,\quad\qquad
\brP^{BD}\delta\Gamma_{BCD}=2\brP_{C}{}^{B}\partial_{B}\delta d+\na^{B}\delta P_{BC}\,,\\
\delta\Gamma_{CAB}P^{CF}\brP^{BG}=(P\na)^{F}(P\delta P\brP)_{A}{}^{G}+(\brP\na)_{A}(P\delta P\brP)^{FG}-(\brP\na)^{G}(P\delta P\brP)^{F}{}_{A}\,,\\
\delta\Gamma_{CAB}\brP^{CF}P^{BG}=(P\na)^{G}(P\delta P\brP)_{A}{}^{F}-(\brP\na)^{F}(P\delta P\brP)^{G}{}_{A}-
(P\na)_{A}(P\delta P\brP)^{GF}\, ,
\ea
\label{usefulPdeltaG}
\ee
we can simplify $\Theta^A$ into the form
\be
\Theta^{A}(d,P,\delta d,\delta P)=4(P-\brP)^{AB}\partial_{B}\delta d-2\na_{B}\delta P^{AB}\,.
\label{Theta}
\ee

Consider now arbitrary generalized diffeomorphism gauge transformations. They are generated by the generalized Lie derivatives, so
\be
\ba{ll}
\ \ \delta_{X}d  \ \ =\ \ \ \hcL_{X}d &=X^{A}\partial_{A}d-\half\partial_{A}X^{A}=-\half\na_{A}X^{A}\,,
\\
\delta_{X}P_{AB}=\hcL_{X}P_{AB}&=X^{C}\partial_{C}P_{AB}+(\partial_{A}X^{C}-\partial^{C}X_{A})P_{CB}
+(\partial_{B}X^{C}-\partial^{C}X_{B})P_{AC}\\
{}&=(\na_{A}X^{C}-\na^{C}X_{A})P_{CB}
+(\na_{B}X^{C}-\na^{C}X_{B})P_{AC}\\
{}&=2(\brP\na)_{(A}(PX)_{B)}-2(P\na)_{(A}(\brP X)_{B)}\,.
\ea
\label{LieDerivatives}
\ee
These equations give the $\ODD$-covariant Killing equations in DFT:
\be
\ba{ll}
 \na_A X^A = 0\,
\qquad \mbox{and} \qquad
 (P\na)_A (\brP X)_B - (\brP\na)_B (PX)_A = 0\,.
\ea
\label{Killing}
\ee
From \eqref{LieDerivatives}, it also follows that the connection transforms as \cite{Jeon:2011cn}
\be
{\delta_{X}\Gamma}_{CAB}=\hcL_{X}\Gamma_{CAB}+2\left[(\cP+\brcP)_{CAB}{}^{FDE}-
\delta_{C}^{~F}\delta_{A}^{~D}\delta_{B}^{~E}\right]\partial_{F}\partial_{[D}X_{E]}\,.
\ee
This is in fact the generic variation~(\ref{deltaGamma}) when restricted to the generalized diffeomorphism gauge transformations.

To derive off-shell conserved Noether current, we start with the covariance of the weight-one Lagrangian under the general diffeomorphism gauge transformation:
\be
\delta_{X}\cL_{\NS}=\partial_{A}\left(X^{A}\cL_{\NS}\right)\,.
\ee
It gives the identity:
\be
\ba{l}
\partial_{A}\left(X^{A}e^{-2d}\cS\right)=
e^{-2d}X_{B}\na_{A}\left[4(P^{AC}\brP^{BD}-\brP^{AC}P^{BD})S_{CD}-\cJ^{AB}\cS\right]\\
\qquad\qquad\quad
+\partial_{A}\left[4e^{-2d}(\brP^{AC}P^{DE}-P^{AC}\brP^{DE})S_{CD}X_{E}+2e^{-2d}(P^{AC}P^{BD}-\brP^{AC}\brP^{BD})\delta_{X}\Gamma_{BCD}\right]\\
\qquad\qquad\quad
+\partial_{A}\left(X^{A}e^{-2d}\cS\right)\,,
\ea
\label{ident0}
\ee
implying that the sum of the first line and the second line should vanish identically. Actually, the above identity~(\ref{ident0}) holds not just for generalized gauge transformations but also for arbitrary local transformations generated by the vector field, $X^{A}$. Therefore, the first line and the second line of (\ref{ident0}) ought to vanish independently.~\footnote{For those readers not convinced, consider the case where the vector field $X^{A}$ is a distribution on an arbitrary point as a Dirac delta function and integrate (\ref{ident0}) over a section. This will confirm that the first line vanishes by itself.} Consequently, we obtain an off-shell, covariantly conserved two-index curvature field $G_{AB}$, which we propose as the DFT counterpart of the  \textit{Einstein curvature tensor}:
\be
\ba{ll}
G^{AB}:=2(P^{AC}\brP^{BD}-\brP^{AC}P^{BD})S_{CD}-\half\cJ^{AB}\cS\,
\quad \mbox{obeying} \quad
\na_{A}G^{AB}=0\,,
\ea
\label{deftwoindexJ}
\ee
and  an \textit{off-shell, covariantly conserved  Noether current},
\be
\ba{l}
J^{A}:=4(\brP^{AC}P^{D}{}_{E}-P^{AC}\brP^{D}{}_{E})X^{E}S_{CD}+2(P^{AC}P^{BD}-\brP^{AC}\brP^{BD})\delta_{X}\Gamma_{BCD}\,,\\
\mbox{obeying} \quad \na_{A}J^{A}=0\,,\qquad\partial_{A}(e^{-2d}J^{A})=0\,.
\ea
\label{defJ}
\ee

We can further decompose the conservation law of the generalized Einstein curvature $G_{AB}$ by projecting
$\na_{A}G^{AB}=0$ with $P$ or $\brP$. We obtain a pair of   conservation relations:
\be
\ba{ll}
\na_{A}\left(4P^{AC}\brP^{BD}S_{CD}- \brP^{AB}\cS\right)=0\,,
\quad&\quad
\na_{A}\left(4\brP^{AC}P^{BD}S_{CD}+ P^{AB}\cS\right)=0\,,
\ea
\label{twofoldconserv}
\ee
which can be re-expressed as
\be
\ba{ll}
4(P\na)_{A}(PS\brP)^{AB}-(\brP\na)^{B}\cS=0\,,
\quad&\quad
4(\brP\na)_{A}(\brP SP)^{AB}+ (P\na)^{B}\cS=0\,.
\ea
\ee
We see that these conservation relations are precisely the `differential Bianchi identities' obtained  previously  in \cite{Hohm:2011si}. While the difference of the two projected curvatures in (\ref{twofoldconserv}) yields back the generalized Einstein tensor~(\ref{deftwoindexJ}), their sum leads to nothing new. Rather, it yields the conservation relation of `symmetric' Einstein curvature tensor:
\be
\ba{ll}
\na_{A}\left[\cG^{(AB)} \right]=0 \quad \mbox{where}
\quad
\cG^{(AB)} := G^{AC}\cH_{C}{}^{B}=-4(PS\brP)^{(AB)}-\half\cH^{AB}\cS
\,.
\ea
\label{defsymEinstein}
\ee
Note also that the equations of motion~(\ref{EOM}) are equivalent to the vanishing of the generalized Einstein curvature,
\be
G^{AB}\simeq0\,.
\ee

Note that the off-shell, conserved Noether current is expressible as
\be
J^{A}=8(\brP SP)^{[AB]}X_{B} +4\na_{B}\left[  (P\na)^{(A}(\brP X)^{B)}-
(\brP\na)^{(A}(PX)^{B)}-\half\cH^{AB}\na_{C}X^{C}\right]\,,
\ee
and, also in terms of the Einstein curvature  tensor and the $\Theta$-term~(\ref{Theta}), as
\be
J^{A}=-2G^{AB}X_{B}+\Theta^{A}(d,P,\delta_{X}d,\delta_{X}P)-\cS X^{A}\,.
\label{JGTheta}
\ee
This then leads to the on-shell Noether current:
\be
H^{A}:=\Theta^{A}(d,P,\delta_{X}d,\delta_{X}P)-\cS X^{A}\,.
\label{defH}
\ee
Taking the divergence, we obtain
\be
\na_{A}H^{A}=2G^{AB}\na_{A}X_{B}=2\cS\delta_{X}d
-4(PS\brP)^{AB}\delta_{X}P_{AB}\simeq 0\,.
\ee
Indeed, the right-hand side vanishes either on-shell or, alternatively, for a Killing vector, $X^A$,  satisfying (\ref{Killing}). 
We would like to further re-express the Noether current such that conservation relation is manifest. We do so by
searching for a skew-symmetric \textit{Noether potential}, $K^{AB}$, in terms of which the off-shell conserved Noether current is given by
\be
\ba{ll}
 e^{-2d}J^{A}=\partial_{B}(e^{-2d}K^{AB})+\Phi\partial^{A}\Phi^{\prime}\,,\quad&\quad
 K^{AB}=-K^{BA}\,.
\ea
\ee
In this form, the conservation relation is manifest up to the section condition. Note that $\Phi\partial^{A}\Phi^{\prime}$ takes the generic form of a `derivative-index-valued vector'~\cite{Park:2013mpa} which generates the coordinate gauge symmetry~(\ref{aTensorCGS}). As such, upon imposing the section condition, it is automatically conserved.
Moreover, it will not contribute to the global charge in the next subsection, which is defined on a given choice of the section by the spatial integral of the conserved charge density. Hence, we may freely drop off such derivative-index-valued vectors and take the Noether current up to the  derivative-index-valued-vectors as
\be
e^{-2d}J^{A}\equiv\partial_{B}(e^{-2d}K^{AB})\,.
\ee

To find explicit expression of the Noether potential, we utilize the projection field identities (\ref{identitiesS}) and commutator relations:
\be
\ba{ll}
\left[(P\na)_{B},(\brP\na)_{A}\right]X^{B}=(\brP SP)_{AB}X^{B}\,,
\quad&\quad
\left[(\brP\na)_{B},(P\na)_{A}\right]X^{B}=(P S\brP)_{AB}X^{B}\,,
\ea
\label{usefulcommutator}
\ee
and rewrite the off-shell conserved current $J^A$ in the form
\be
J^{A}=4\na_{B}\left[(\brP\na)^{[A}(PX)^{B]}-(P\na)^{[A}(\brP X)^{B]}\right]
+2\partial^{A}\left[(\brP-P)^{BC}\na_{B}X_{C}\right]\,.
\label{bareJ}
\ee
This expression naturally suggests to define the skew-symmetric Noether potential as
\be
K^{AB}:=4(\brP\na)^{[A}(PX)^{B]}-4(P\na)^{[A}(\brP X)^{B]}\,.
\label{Kpotential}
\ee
With this definition, we finally get
\be
\ba{ll}
e^{-2d}J^{A}&=\partial_{B}(e^{-2d}K^{[AB]})+2 e^{-2d}\partial^{A}\left[(\brP-P)^{BC}
\na_{B}X_{C}\right]\\
{}&+2 e^{-2d}(P\partial^{A}P\brP)^{BC}\left[
(\brP\na)_{C}(PX)_{B}+(P\na)_{B}(\brP X)_{C}\right]\,.
\ea
\label{edJ}
\ee
In going from (\ref{bareJ}) to (\ref{edJ}), one needs to take care of the semi-covariant derivative connections. It turns out that they merely yield derivative-index-valued vectors, as in the second line of (\ref{edJ}). For this, it is worth to note that  $\Gamma^{A}{}_{BC}P^{B}{}_{D}\brP^{C}{}_{E}=(P\partial^{A}P\brP)_{DE}$ is a derivative-index-valued vector as well.

\subsection{Conserved  Global  Charges}
We now proceed to construct conserved global charges, which will constitute generators of asymptotic symmetry algebra. The starting point is the two-form Komar function. While our Noether potential~(\ref{Kpotential}) can correctly reproduce the two-form Komar integrand, it is well known that the Komar integrand itself needs to be further corrected~\cite{Wald:1993nt,Iyer:1994ys,Iyer:1995kg} (see also  \cite{Kim:2013zha} and references therein).  Here, we derive such a correction  and  define a corresponding   conserved global charge.

We start by rewriting the semi-covariant four-index Riemann curvature in terms of the semi-covariant derivative acting on the connection:
\be
\ba{ll}
S_{ABCD}=& \half\left(
\Gamma^{E}{}_{AB}\Gamma_{ECD}+\Gamma_{CA}{}^{E}\Gamma_{DBE}-
\Gamma_{CB}{}^{E}\Gamma_{DAE}+\Gamma_{AC}{}^{E}\Gamma_{BDE}-
\Gamma_{AD}{}^{E}\Gamma_{BCE}\right)\\
{}&+\na_{[A}\Gamma_{B]CD}+\na_{[C}\Gamma_{D]AB}\,,
\ea
\ee
This enables us to isolate the two-derivative terms (`accelerations') from the one-derivative terms (`velocities') in the DFT Lagrangian:
\be
\ba{ll}
& e^{-2d}(P^{AC}P^{BD}-\brP^{AC}\brP^{BD})S_{ABCD}\\
= & e^{-2d}(P^{AC}P^{BD}-\brP^{AC}\brP^{BD})
(\Gamma_{AC}{}^{E}\Gamma_{BDE}-\Gamma_{AB}{}^{E}\Gamma_{DCE}+\half\Gamma^{E}{}_{AB}\Gamma_{ECD})\\
& + 2\partial_{A}\left[
e^{-2d}(P^{AC}P^{BD}-\brP^{AC}\brP^{BD})
\Gamma_{BCD}\right]\\
= &
e^{-2d}(P^{AC}P^{BD}-\brP^{AC}\brP^{BD})
(\Gamma_{AC}{}^{E}\Gamma_{BDE}-\Gamma_{AB}{}^{E}\Gamma_{DCE}+\half\Gamma^{E}{}_{AB}\Gamma_{ECD})\\
& +\partial_{A}\left[
e^{-2d}\Big\{4(P-\brP)^{AB}\partial_{B}d -2\partial_{B}P^{AB}\Big\}\right]\,.
\ea
\ee
Motivated by this observation, we define a composite vector field:
\be
\ba{ll}
B^{A}&:=2(P^{AC}P^{BD}-\brP^{AC}\brP^{BD})
\Gamma_{BCD} \\
& =4(P-\brP)^{AB}\partial_{B}d -2\partial_{B}P^{AB}\,.
\ea
\label{defB}
\ee
Note that this is not quite a diffeomorphism covariant vector: because of (\ref{diffGamma}), $B^A$ transforms   anomalously,
\be
\delta_{X}B^{A}=\hcL_{X}B^{A}+4(\brP^{AC}\brP^{BD}-P^{AC}P^{BD})\partial_{B}\partial_{[C}X_{D]}\,.
\label{Banomaly}
\ee
Only if the vector field $X^A$ can be restricted to satisfy $\partial_B \partial_{[C} X_{D]}=0$, the composite vector field $B^A$ transforms covariantly under the generalized diffeomorphism gauge transformations. We will see momentarily that this condition can be arranged by modifying the Noether potential in a specific way.

The idea is that we would like to remove the two-derivative terms. To do so, we consider modifying the DFT Lagrangian with a specific surface term:
\be
\ba{ll}
\widehat{\cL}_{\NS}&=\cL_{\NS}-\partial_{A}(e^{-2d}B^{A})\\
{}&=e^{-2d}(P^{AC}P^{BD}-\brP^{AC}\brP^{BD})
(\Gamma_{AC}{}^{E}\Gamma_{BDE}-\Gamma_{AB}{}^{E}\Gamma_{DCE}+\half\Gamma^{E}{}_{AB}\Gamma_{ECD})\,.
\ea
\ee
The idea is analogous to the modification of the Einstein-Hilbert action to the Schr\"odinger action \textit{a la} Dirac~\cite{DiracGRBook} that is free of two-derivative terms. While the equations of motion remain intact, the theta term in the variation of the Lagrangian, (\ref{varL}), gets modified to
\be
e^{-2d}\widehat{\Theta}^{A}(d,P,\delta d,\delta P)=e^{-2d}\Theta^{A}(d,P,\delta d,\delta P)-\delta (e^{-2d}B^{A})\,,
\label{Thetapr}
\ee
such that the new theta term, $\widehat{\Theta}^{A}(d,P,\delta d,\delta P)$, no longer  contains the derivative of the variations. In particular, for the generalized diffeomorphism gauge transformations, we have from (\ref{Banomaly}) that
\be
\ba{ll}
e^{-2d}\widehat{\Theta}^{A}(d,P,\delta_{X}d,\delta_{X}P)
{}=& e^{-2d}\Theta^{A}(d,P,\delta_{X}d,\delta_{X}P)+2\partial_{B}(e^{-2d}X^{[A}B^{B]})-e^{-2d}B_{B}\partial^{A}X^{B}\\
{}&-X^{A}\partial_{B}(e^{-2d}B^{B})+4e^{-2d}(P^{AC}P^{BD}-\brP^{AC}\brP^{BD})\partial_{B}\partial_{[C}X_{D]}\,.
\ea
\ee
The off-shell Noether current~(\ref{JGTheta}) now receives extra contributions
\be
\ba{ll}
e^{-2d}\widehat{J}^{A} & = e^{-2d}\left[-2G^{AB}X_{B}+\widehat{\Theta}^{A}(d,P,\delta_{X}d,\delta_{X}P)-(\cS-\na_{B}B^{B})X^{A}\right] \\
& = e^{-2d}J^{A} \\
  &\ \ \ +2\partial_{B}(e^{-2d}X^{[A}B^{B]})-e^{-2d}B_{B}\partial^{A}X^{B}
+4e^{-2d}(P^{AC}P^{BD}-\brP^{AC}\brP^{BD})\partial_{B}\partial_{[C}X_{D]}\,.
\ea
\label{Jpr}
\ee
Correspondingly, we have the modified Noether potential
\be
\widehat{K}^{AB}=K^{AB}+2X^{[A}B^{B]}\,.
\label{Kpr}
\ee

We need to ensure that the modified Noether current is conserved. Taking the divergence, one finds that
\be
\partial_{A}(e^{-2d}\widehat{J}^{A})=4e^{-2d}(P^{AC}P^{BD}-\brP^{AC}\brP^{BD})\na_{[A}(\partial_{B]}\partial_{[C}X_{D]})\,.
\ee
Thus, the modified off-shell Noether current is not always conserved. However, we can ensure the conservation relation provided we impose the diffeomorphism vector field to obey the condition
\be
\na_{[A}(\partial_{B]}\partial_{[C}X_{D]})=0\,,
\ee
or more strongly the condition
\be
\partial_{B}\partial_{[C}X_{D]}=0\,.
\label{ASSUMP}
\ee
Wonderfully, the latter condition is precisely the condition we needed in order to ensure the composite vector field $B^{A}$ transform covariantly~(\ref{Banomaly}). The simplest example of such restricted vector field is when $X^{A}$ is a constant vector, corresponding to a rigid translation in doubled spacetime.

With the conserved modified Noether current at hand, we  finally obtain the conserved global charge as surface integral:
\be
\boxed{
Q[X]:=\int_{\cM}\rmd^{{\scriptscriptstyle{D{-1}}}} x_{A} ~e^{-2d}\widehat{J}^{ A}=\oint_{\partial\cM}\rmd^{{\scriptscriptstyle{D{-2}}}}x_{AB}~e^{-2d}\left(K^{AB}+2X^{[A}B^{B]}\right)\,
} \, .
\label{defQ}
\ee
Here, $\cM$ denotes a timelike hypersurface inside the section, $\Sigma_{D} =\mathbb{R}_{t}\times\cM$, while $\partial\cM$ corresponds to its asymptotic boundary.  \\

Intuitively, we can also motivate the conserved global charge proposed above from the method adopted by Wald~\cite{Wald:1993nt,Iyer:1994ys,Iyer:1995kg}.
Modulo the equations of motion, the variation of the on-shell conserved Noether current~(\ref{defH}) reads
\be
\delta(e^{-2d}H^{A})\simeq e^{-2d}\Omega^{A}(\delta,\delta_{X})
-2\partial_{B}\left(e^{-2d}X^{[A}\Theta^{B]}\right)
+e^{-2d}\Theta_{B}\partial^{A}X^{B}\,,
\label{varH}
\ee
where $\Omega^{A}$ denotes the (Hamiltonian) symplectic structure defined by
\be
e^{-2d}\Omega^{A}(\delta_{1},\delta_{2}):=\delta_{1}\left[e^{-2d}\Theta^{A}(d,P,\delta_{2}d,\delta_{2}P)\right]
-\delta_{2}\left[e^{-2d}\Theta^{A}(d,P,\delta_{1}d,\delta_{1}P)\right]\,.
\ee
The above variation~(\ref{varH}) then reveals an on-shell relation
\be
e^{-2d}\Omega^{A}(\delta,\delta_{X})\simeq
\partial_{B}\left[\delta\left(e^{-2d}K^{AB}\right)
+2\left(e^{-2d}X^{[A}\Theta^{B]}\right)\right]\,-e^{-2d}\Theta_{B}\partial^{A}X^{B}\,.
\ee
Again, the last term is a derivative-index-valued vector and can be dropped off when integrated over a section.   Finally, by assuming proper asymptotic fall-off behaviour at infinity, the left-hand side of (\ref{Thetapr}) can be  made to vanish at infinity~\footnote{See Appendix~\ref{APPDEMON} for explicit demonstration of this for the asymptotically flat hypersurface.}. This facilitates to approximate
\be
e^{-2d}\Omega^{A}(\delta,\delta_{X})\approx
\partial_{B}\delta\left[e^{-2d}\left(K^{AB}
+2X^{[A}B^{B]}\right)\right]\,.
\label{Approx}
\ee
The final expression then supports the validity of our proposed expression for the conserved global  charge~(\ref{defQ}).

\subsection{Extension to Yang-Mills and Cosmological Constant Sectors}
A proper account of low-energy string theory requires inclusion of the Yang-Mills sector and the cosmological constant in addition to the NS-NS sector.
Here, we consider the DFT in which the NS-NS sector is coupled to Yang-Mills sector and the cosmological constant is included~(\ref{cL}),
\be
\cL_{\DFT}=e^{-2d}\left[\cS-2\Lambda+\gYM^{-2}\Tr\left(P^{AC}\brP^{BD}\cF_{AB}\cF_{CD}\right)\right]\, ,
\ee
and construct corresponding extension of the conserved global charges.

Consider arbitrary variations of the projection field and the vector potential,
\be
\ba{l}
\delta(\cF_{AB})=2\cD_{[A}\delta V_{B]}-\delta\Gamma^{C}{}_{AB}V_{C}\,,\qquad\cD_{A}:=\na_{A}-i\left[V_{A}\,,\,~~\right]\,,\\
P_{A}{}^{C}\brP_{B}{}^{D}\delta(\cF_{CD})=2
P_{A}{}^{[C}\brP_{B}{}^{D]}\cD_{C}\delta V_{D}-\na^{C}(P\delta P\brP)_{AB}V_{C}\,.
\ea
\ee
This induces variation of the YM part in the DFT Lagrangian as
\be
\ba{ll}
\delta\Tr\left(P^{AC}\brP^{BD}\cF_{AB}\cF_{CD}\right)=&
-4\Tr\left[\delta V_{B}\cD_{A}\left(P\cF\brP\right)^{[AB]}\right]\\
{}&+2(P\delta P\brP)^{AB}\Tr\left[(P\cF\cH\cF\brP)_{AB}
+\na^{C}\left\{(P\cF\brP)_{AB}V_{C}\right\}\right]\\
{}&+\na_{A}\Tr\left[4(P\cF\brP)^{[AB]}\delta V_{B}-2V^{A}(P\cF\brP)_{CD}\delta P^{CD}\right]\,.
\ea
\label{YMvar}
\ee
For local variations, from the first line, we find the YM equation of motion:~\footnote{
It is useful to note
\be
(P\cF\brP)_{[AB]}=(\brP\cF P)_{[AB]}\,.
\ee
}
\be
\cD_{A}\left(P\cF\brP\right)^{[AB]}\simeq 0\,.
\label{YMEOM}
\ee
From the second line in (\ref{YMvar}), we find that the  equation of motion of the projection field changes  from (\ref{EOM}) to
\be
(PS\brP)_{AB}+\half\gYM^{-2}\left[(P\cF\cH\cF\brP)_{AB}+\na_{C}\left\{(P\cF\brP)_{AB}V^{C}\right\}\right]\simeq0\,.
\ee
We also recall that the equation of motion of the dilaton is now modified to
\be
\cS-2\Lambda+\gYM^{-2}\Tr\left(P^{AC}\brP^{BD}\cF_{AB}\cF_{CD}\right)\simeq 0\,.
\ee
Once again, the DFT Lagrangian vanishes on-shell. For consistency, from (\ref{YMdiffanomaly})   and  (\ref{YMgauge}),   it is straightforward to check that
\be
(P\cF\cH\cF\brP)_{AB}+\na_{C}\left[(P\cF\brP)_{AB}V^{C}\right]\,.
\label{newcovF}
\ee
is indeed fully covariant  under both the generalized diffeomorphisms and the YM gauge transformations.

The last line  in (\ref{YMvar}) is the YM contribution to $\Theta^{A}$.  Then, in steps completely parallel to the analysis of the NS-NS sector DFT as carried out in section~\ref{SECNSNS}, we can straightforwardly obtain the off-shell conserved Noether current associated with the diffeomorphism transformation to the total DFT.  Modulo the part identifiable with derivative-index-valued-vectors, we get
\be
\ba{ll}
e^{-2d}\widehat{J}_{\rm total}^{A}&\equiv e^{-2d}\widehat{J}^{A}+{12}\gYM^{-2}\partial_{B}\Tr\left( e^{-2d}(P\cF\brP)^{[AB}V^{C]}X_{C}\right)\\
{}& = \partial_{B}\left[e^{-2d}K^{[AB]}+{12} \gYM^{-2}e^{-2d}\Tr\left\{(P\cF\brP)^{[AB}V^{C]}X_{C}\right\}\right]\,.
\ea
\ee
The conserved global charge is now generalized to
\be
\boxed{
Q_{\rm total} [X]=\oint_{\partial\cM}\rmd x_{AB}~e^{-2d}\left[K^{[AB]}+2X^{[A}B^{B]}
+{1 \over \gYM^{2}}\Tr\left\{12 (P\cF\brP)^{[AB}V^{C]}X_{C}\right\}\right]\,} \, .
\label{MAINFINAL}
\ee
It is worth to note that the cosmological constant term does not contribute to the global charges.



\section{Applications\label{SECAPP}}

In this section, we apply the general formula of the conserved global charge~\eqref{defQ} to various asymptotically flat string backgrounds.
In sections \ref{sec:null-wave} and \ref{sec:F1-string},
we consider the null wave solutions in DFT \cite{Berkeley:2014nza}
and calculate their ADM $2D$-momenta,
which are the conserved global charges associated with the global translations.
In section \ref{sec:Non-Riemannian}, by  performing  further dualities,
we discuss the $2D$-momentum for a non-Riemannian background reported in \cite{Lee:2013hma}.
In section \ref{sec:RN-BH}, as a demonstration for the YM-coupled DFT,
we consider the Reissner-Nordstr\"om black hole.
In section \ref{sec:5-brane}, we consider the background of black 5-branes.
Finally, in section \ref{sec:linear-dilaton},
as an application to a non-asymptotically flat background,
we consider a linear dilaton background and show that
the known result can be correctly reproduced
by introducing an extra counterterm to the boundary.

Henceforth,  we fix the $D$-dimensional section, $\Sigma_{D}$,
to be independent of the winding coordinates, $\tilde{x}_\mu$,
as in \eqref{asection}.
Further, we decompose the generalized metric into a constant part and a fall-off part,
\begin{align}
 \cH_{AB} = \cH^{(0)}_{AB} + \Delta_{AB} \,,\quad
 \cH^{(0)}_{AB}:= \begin{pmatrix} \eta^{\mu\nu} & -\eta^{\mu\rho}\,b_{\rho\nu} \cr b_{\mu\rho}\,\eta^{\rho\nu} & \eta_{\mu\nu} \end{pmatrix} \,,
\end{align}
where $\eta_{\mu\nu}=\mathrm{diag}(-1,1,\dotsc,1)=\eta^{\mu\nu}$ is the flat Minkowski metric.
The asymptotic values of the $B$-field is denoted by $b_{\mu\nu}$ and
it is assumed to be constant.
In this paper, we consider two kinds of asymptotically flat backgrounds:
(1) backgrounds with the topology, $\cM = \mathbb{R}^{D-1}$,
which include the (higher-dimensional) Schwarzschild solution;
and
(2) backgrounds with $\cM = T^p \times \mathbb{R}^{D-p-1}$,
where $T^p$ is a $p$-torus,
which include backgrounds of $p$-branes wrapped on the $p$-torus.
For each background, we introduce the coordinates for the $D$-dimensional section, $\Sigma_D$, as
\begin{align}
 (1)\,:&\quad (x^\mu) = (t,\,x^i)\,,\quad i=1,\dotsc,D-1\,,
\label{eq:BG1}
\\
 (2)\,:&\quad (x^\mu) = (t,\,z^s,\,y^m)\,,\quad s=1,\dotsc,p \,, \quad m=1,\dotsc,D-p-1 \,,
\label{eq:BG2}
\end{align}
and define the radius respectively by
\begin{align}
 (1)\,:\quad r:= \sqrt{\delta_{ij}\,x^i\,x^j} \,,\qquad
 (2)\,:\quad r:= \sqrt{\delta_{mn}\,y^m\,y^n} \,.
\end{align}
Then, in terms of the radius, we assume a simple fall-off behaviour,
\begin{align}
 \Delta_{AB}=\mathcal{O}(r^\alpha)\,,\quad
 e^{-2d}=1+\mathcal{O}(r^\alpha) \,,\quad
 \alpha<0 \,.
\label{eq:fall-off}
\end{align}
We have $\alpha = - (D-3)$ for the Schwarzschild solution
while $\alpha = - (D-p-3)$ for $p$-brane solutions.

Before considering examples, we make remarks on the conserved global charge defined in \eqref{defQ}.
The  charge consists of an integral of $K^{AB}$ and $2X^{[A}B^{B]}$.
The explicit form of the DFT-Noether potential
is given in Appendix \ref{app:Noether}, and its $\mu\nu$ components
with a Riemannian parametrization \eqref{aPARAN} become
\begin{align}
 K^{\mu\nu}[X] = 2\, \xi^{[\mu;\nu]} - H^{\mu\nu\rho}\, \zeta_\rho \,,
\label{eq:Komar-munu}
\end{align}
where $\xi^{\mu;\nu}$ is the conventional Riemannian covariant derivative and $X^A$ is parametrized in a manner parallel to (\ref{YMpara}) by
\begin{align}
 X^A= \begin{pmatrix}
 \zeta_\mu +B_{\mu\nu}\,\xi^\nu \cr \xi^\mu
 \end{pmatrix} \,.
\label{eq:X-param}
\end{align}
As this parametrization suggests, in this paper, we define the ADM $2D$-momentum as
\begin{align}
 P_A := Q[E_A]\,,\quad
 E_A := \begin{pmatrix} \tilde{\partial}^\mu \cr \partial_\mu - b_{\mu\nu}\,\tilde{\partial}^\nu \end{pmatrix} \,,
\label{eq:ADM-def}
\end{align}
namely, the conserved global charge for $X$ with constant $\xi^\mu$ or $\zeta_\mu$. We also replace $B_{\mu\nu}$ by $b_{\mu\nu}$ since the global charge is evaluated
as the surface integral at infinity.
In section \ref{sec:F1-string}, we discuss the importance of this definition of the ADM momentum.

A remark is in order. DFT is manifestly covariant under O(D,D) rotations which act on both the tensor indices and the arguments of the tensor, \textit{i.e.} coordinates. In this case, the  whole configuration including the section itself  is  rotated, and there should be no change in physics. However, if  a specific given background admits an isometry, there is ambiguity of choosing the section. Without rotating the section, it is possible to rotate the tensor indices only and this can generate a physically different configuration.  Thus,  while our global charge is manifestly O(D,D) covariant, it may not transform covariantly if we keep the section fixed and rotate only the tensor indices.

\subsection{Pure Einstein gravity}
\label{eq:pure-Einstein}

Here, considering the pure Einstein gravity, 
\textit{i.e.}
\begin{align}
 \Phi = 0\,,\qquad\qquad B_{\mu\nu} =0\,,
\end{align}
we show that the ADM mass, $P_t$, defined on \eqref{eq:ADM-def},
evaluated for a background with $\cM = \mathbb{R}^{D-1}$,
correctly reproduces the well-known ADM mass formula \cite{Arnowitt:1959ah},
\begin{align}
 E_{\mathrm{ADM}} = \frac{1}{2\kappa_{\scriptscriptstyle{D}}^2}\,\int_{S^{D-2}_\infty} \rmd\Omega_{{\scriptscriptstyle{D{-2}}}} \,r^{D-2}\,\hat{n}^k\,G^{ij}\,\bigl(\partial_i G_{kj} - \partial_k G_{ij}\bigr)\,,
\label{eq:ADM-mass}
\end{align}
where $\hat{n}^k\partial_k = \partial_r$ is a radial vector that becomes a unit vector at infinity.

From $B_{\mu\nu}=0$, the DFT-Noether potential \eqref{eq:Komar-munu} is reduced to the standard Komar potential:
\begin{align}
 K^{\mu\nu}[X] = 2\, \xi^{[\mu;\nu]} \,.
\end{align}
On the other hand, $B^\mu$, in the Cartesian coordinates, simply becomes
\begin{align}
 B^\mu &= -2\,\cH^{\mu\nu}\,\partial_\nu\ln\sqrt{|G|} - \partial_\nu \cH^{\mu\nu}
 = G^{\mu\nu}\,G^{\rho\sigma}\, \bigl(\partial_{\rho} G_{\nu\sigma}-\partial_{\nu} G_{\rho\sigma}\bigr) \,.
\end{align}
In particular, the identically conserved current, $\partial_{\nu}(\delta^{[\mu}{}_{\lambda}B^{\nu]})$, corresponds to the Einstein pseudo-tensor \textit{a la} Dirac~\cite{DiracGRBook}.

Our definition of the ADM mass is now
\begin{align}
 P_t&=Q[\partial_t] = \int_{\partial\cM}\rmd^{{\scriptscriptstyle{D{-2}}}}x_{\mu\nu}\,\sqrt{|G|}\,
        \bigl(K^{\mu\nu}[\partial_t] + 2X^{[\mu}\,B^{\nu]}\bigr)
\nonumber\\
 &= 2\int_{S^{D-2}_\infty}\rmd^{{\scriptscriptstyle{D{-2}}}}x_{tr}\, \sqrt{|G|}\,\bigl(K^{tr}[\partial_t] + B^r \bigr)\,,
\end{align}
and we have as $r$ goes to infinity,
\begin{align}
 K^{tr}[\partial_t] &= -2\,G^{t\mu}\,G^{r\nu}\,\partial_{[\mu} G_{\nu]t}
  \approx -\hat{n}^k\,\bigl(\partial_k G_{tt} - \partial_t G_{kt}\bigr) \,,
\\
 B^r &\approx \hat{n}^k\,\bigl(\partial_k G_{tt}-\partial_t G_{kt}\bigr)
 +\hat{n}^k\,G^{ij}\,\bigl(\partial_i G_{kj} - \partial_k G_{ij}\bigr) \,.
\end{align}
These precisely coincide with the known results of \cite{Iyer:1994ys}. Finally, summing them up and using
the expression for the integral measure at the spatial infinity, we have
\begin{align}
 2\int_{S^{D-2}_\infty}\rmd^{{\scriptscriptstyle{D{-2}}}}x_{tr} \sqrt{|G|}\quad\cdots\quad
 = \int_{S^{D-2}_\infty} \rmd \Omega_{{\scriptscriptstyle{D{-2}}}} \, r^{D-2} \quad \cdots \, . 
\end{align}
Recalling that we are
choosing a unit $2\kappa_{\scriptscriptstyle{D}}^2= 1 (=16\pi G_N)$,
we thus obtain the ADM mass formula \eqref{eq:ADM-mass}.

\subsection{Null Wave}
\label{sec:null-wave}

We here consider the null-wave solution of DFT \cite{Berkeley:2014nza}
\begin{align}
 \rmd s^2 &= \eta_{\mu\nu}\,\rmd x^\mu\,\rmd x^\nu + \eta^{\mu\nu}\,\rmd \tilde{x}_\mu\,\rmd \tilde{x}_\nu
  + (H-1)\,\bigl[(\rmd t-\rmd z)^2 - (\rmd \tilde{t}+\rmd \tilde{z})^2\bigr] \,,
\\
 e^{-2d} &= 1 \,,\quad
 H(r) = 1 + \frac{\gamma_{\scriptscriptstyle{D}}}{r^{D-4}}  \,,
\end{align}
where $\gamma_{\scriptscriptstyle{D}}$ is a certain constant which depends on the dimension, $D$.
This is another purely gravitational solution in $D$-dimensions:
\begin{align}
 \rmd s^2 &= (H-2)\,\rmd t^2 -2\,(H-1)\,\rmd t\,\rmd z + H\,\rmd z^2 + \delta_{mn}\,\rmd y^m\,\rmd y^n \,,
\\
 \Phi &= 0 \,,\qquad B_{\mu\nu} = 0 \,.
\end{align}
Here, we compactified the $z$-direction with a radius $R_z$, in order to make the value of the global charge finite. The remaining $y^m$ directions are treated as non-compact.

In this background, using $\partial_{\mu} d=0$ and $K^{\mu\nu}=- 2\,G^{\rho[\mu}\,G^{\nu]\delta}\,\partial_{\rho} \xi_{\delta}$, we have at infinity,
\begin{align}
 K^{tr}[\partial_t] \approx -\partial_r H(r) \,,
\quad
 K^{tr}[\partial_z] \approx \partial_r H(r) \,,
\quad
 B^r\approx 0 \,.
\end{align}
Then, it is straightforward to show that the ADM energy and the momentum in the $z$-direction becomes
\begin{align}
 P_t&= Q[\partial_t]
 = 2\,\int_{\partial\cM}\rmd^{{\scriptscriptstyle{D{-2}}}}x_{tr}\, K^{tr}[\partial_t]
  = -\int\rmd z \int_{S^{D-3}_\infty} \rmd\Omega_{{\scriptscriptstyle{D{-3}}}} \,R^{D-3}\,\partial_r H
\nonumber\\
 &= (D-4)\,\gamma_{\scriptscriptstyle{D}}\,(2\pi R_z)\, \Omega_{{\scriptscriptstyle{D{-3}}}} \,,
\\
 P_z&= Q[\partial_z] = 2\,\int_{\partial\cM}\rmd^{{\scriptscriptstyle{D{-1}}}}x_{tr}\, K^{tr}[\partial_z]
  = \int\rmd z \int_{S^{D-3}_\infty} \rmd\Omega_{{\scriptscriptstyle{D{-3}}}} \,R^{D-3}\,\partial_r H
\nonumber\\
 &= - (D-4)\, \gamma_{\scriptscriptstyle{D}}\,(2\pi R_z)\, \Omega_{{\scriptscriptstyle{D{-3}}}} \,.
\end{align}
Here, $\partial\cM$ is the surface of constant $t$ and $r{=R}$ in the  $R\to \infty$ limit,
and $\Omega_{{\scriptscriptstyle{D{-3}}}}$ is a surface area of a $(D-3)$-sphere with a unit radius; $\Omega_{{\scriptscriptstyle{D{-3}}}} = 2\pi^{(D-2)/2}/\Gamma((D-2)/2)$.
As the momenta in other directions are trivial, the $2D$-momentum becomes
\begin{align}
 &P^{(\mathrm{wave})}_A
 = (\tilde{P}^t,\, \tilde{P}^z,\, \tilde{P}^m\,;\,P_t,\, P_z,\, P_m) = n_{\scriptscriptstyle{D}}\,(0,\dotsc,0\,;\,+1,-1,0,\dotsc,0) \,,
\\
 &n_{\scriptscriptstyle{D}}:= (D-4)\, \gamma_{\scriptscriptstyle{D}}\,(2\pi R_z)\, \Omega_{{\scriptscriptstyle{D{-3}}}} \,,
\end{align}
which is indeed a null vector at the flat spatial infinity:
\begin{align}
 \bigl(\cH^{(0)}\bigr)^{AB}\,P^{(\mathrm{wave})}_A\,P^{(\mathrm{wave})}_B = 0 \,.
\end{align}
For $D=10$, in our unit of $2\kappa_{10} =1$, the constants become
\begin{align}
 \gamma_{10} = \frac{1}{\pi^3\,(2\pi R_z)^2}\,, \quad \Omega_{7}=\frac{\pi^4}{3} \,,\quad n_{10}= 1/R_z \,.
\end{align}
Namely, the ADM energy and the momentum have just the expected values:
\begin{align}
 P_0 = 1/R_z = -P_z \,.
\end{align}

\subsection{Fundamental String}
\label{sec:F1-string}

As it was pointed out in \cite{Berkeley:2014nza},
the string background can be also constructed by considering a null wave
propagating in a winding direction $\tilde{z}$.
Starting from the doubled wave solution,
by performing a $T$-duality along the $z$ direction,
one can obtain
\begin{align}
 \rmd s^2 &= \eta_{\mu\nu}\,\rmd x^\mu\,\rmd x^\nu + \eta^{\mu\nu}\,\rmd \tilde{x}_\mu\,\rmd \tilde{x}_\nu
  + (H-1)\,\bigl[(\rmd t-\rmd \tilde{z})^2 - (\rmd \tilde{t}+\rmd z)^2\bigr] \,,
\\
 e^{-2d} &= 1 \,,\quad
 H(r) = 1 + \frac{\gamma_{\scriptscriptstyle{D}}}{r^{D-4}} \,.
\label{eq:string-2D}
\end{align}
This corresponds to the following $D$-dimensional Riemannian background:
\begin{align}
 \rmd s^2 &= H^{-1}(r)\,(-\rmd t^2+\rmd z^2) + \delta_{mn}\,\rmd y^m\,\rmd y^n \,,
\\
 B_{tz}&= H^{-1}(r)-1\,,\quad
 \Phi = 0\,,
\end{align}
which  recovers the fundamental string background originally found in \cite{Dabholkar:1990yf},
up to a constant gauge shift in the $B$-field.

For this string background, in a manner similar to the null-wave case, we obtain, at infinity,
\begin{align}
 K^{rt}[\partial_t] \approx -\partial_r H(r) \,,
\quad
 K^{rt}[\tilde{\partial}^z] \approx \partial_r H(r) \,,
\quad
 B^r \approx 0 \,.
\end{align}
Therefore, the nontrivial momenta are
\begin{align}
 P_t&= Q[\partial_t]
  = -\int\rmd z \int_{S^{D-3}_\infty} \rmd\Omega_{{\scriptscriptstyle{D{-3}}}} \,R^{D-3}\,\partial_r H
  = (D-4)\,\gamma_{\scriptscriptstyle{D}}\,(2\pi R_z)\, \Omega_{{\scriptscriptstyle{D{-3}}}} = n_{\scriptscriptstyle{D}} \,,
\\
 \tilde{P}^z&= Q[\tilde{\partial}^z]
  = \int\rmd z \int_{S^{D-3}_\infty} \rmd\Omega_{{\scriptscriptstyle{D{-3}}}} \,R^{D-3}\,\partial_r H
 = -(D-4)\, \gamma_{\scriptscriptstyle{D}}\,(2\pi R_z)\, \Omega_{{\scriptscriptstyle{D{-3}}}} = -n_{\scriptscriptstyle{D}} \,.
\end{align}
As these are the only non-zero components,
the ADM momentum in the string background becomes
\begin{align}
 P^{(\mathrm{string})}_A = n_{\scriptscriptstyle{D}}\,(0,-1,0\dotsc,0\,;\,+1,0,\dotsc,0) \,.
\end{align}
This is the expected result: our global charge formula is covariant under global $\ODD$ transformations, so
the ADM momentum in the string background should be related to that in the null-wave background by an $\ODD$ rotation, $\Lambda_A{}^B$, which corresponds to the $T$-duality along the $z$ direction,
\begin{align}
 P^{(\mathrm{string})}_A = \Lambda_A{}^B\, P^{(\mathrm{wave})}_B \,,\quad
 \tilde{R}_z = \frac{l_s^2}{R_z} \,.
\end{align}
Further, from this duality relation, we identify $n_{10}=1/R_z=(2\pi \tilde{R}_z)\times (2\pi l_s^2)^{-1}$ for $D=10$,
which is the correct known mass for a fundamental string winding in the $z$ direction.

Before proceeding further to the next example,
let us comment on two aspects of the ADM momentum.
\begin{itemize}
\item  \textit{{\textbf{String winding charge}}}.
We can identify the well-known string winding charge (along $z$ direction) given by
\begin{align}
 Q_{\mathrm{F1}} \propto \int_{S^{D-3}_\infty} e^{-2\Phi}\, *_{\scriptscriptstyle{D}} H \,,
\end{align}
as the global charge, $Q[\tilde{\partial}^z]$, for the winding direction $\tilde{z}$, \textit{i.e.~}ADM momentum along the dual direction.
Here, let us focus on a spacetime of toroidal topology,
$\cM= S_z\times \mathbb{R}^{D-2}$,
such that a string is winding along the compactified $z$ direction.
The momentum in the dual $\tilde{z}$ direction reads
\begin{align}
 \tilde{P}^z = Q[\tilde{\partial}^z]
 = 2\,\int_{\partial\cM} \rmd^{{\scriptscriptstyle{D{-2}}}}x_{tr}\, e^{-2\Phi}\,\sqrt{|G|}\,K^{tr}[\tilde{\partial}^z] \,.
\end{align}
By using the formula \eqref{eq:Komar-munu}, the DFT-Noether potential becomes, restricted on the section,
\begin{align}
 K^{\mu\nu}[\tilde{\partial}^z] = -H^{\mu\nu z} \,.
\end{align}
Therefore, the ADM momentum becomes
\begin{align}
 \tilde{P}^z &= Q[\tilde{\partial}^z]
 = -\int\rmd z\int_{S^{D-3}_\infty}\rmd^{{\scriptscriptstyle{D{-3}}}}\theta\,
 \varepsilon_{trz\theta_1\cdots \theta_{D-3}}\,e^{-2\Phi}\,
 \sqrt{|G|}\,G^{t\rho}\,G^{r\sigma}\,G^{z\delta}\,H_{\rho\sigma\delta}
\nonumber\\
 &= 2\pi R_z\,\int_{S^{D-3}_\infty} \rmd^{{\scriptscriptstyle{D{-3}}}}\theta\,e^{-2\Phi}\,\sqrt{|G|}\,
 \varepsilon_{tzr\theta_1\cdots \theta_{D-3}}\, G^{t\mu}\, G^{z\nu}\, G^{r\rho} \, H_{\mu\nu\rho}
\nonumber\\
 &= 2\pi R_z \,\int_{S^{D-3}_\infty} \rmd \Omega_{{\scriptscriptstyle{D{-3}}}}\, e^{-2\Phi}\, *_{\scriptscriptstyle{D}} H \,,
\end{align}
where $\theta_a$ ($a=1,\dotsc, D-3$) are angular coordinates. Namely, the well-known flux integral for the string winding charge (in the $z$ direction)
precisely matches with the quasi-local ADM momentum in the $\tilde{z}$-direction.
This supports the idea \cite{Berkeley:2014nza} that strings are waves in doubled spacetimes.

\item \textit{{\textbf{Physical ADM momentum}}}.
Let us consider a constant shift of the $B$-field in the string background \eqref{eq:string-2D}; $B_{tz}\to B_{tz}+1+c$ where $c$ is a constant.
After this $\ODD$ rotation to a new solution, we can show that the conserved global charges, $Q[\partial_A]$, become
\begin{align}
 Q[\partial_A] = n_{\scriptscriptstyle{D}}\,(0,-1,0\dotsc,0\,;\,-c,0,\dotsc,0) \,.
\end{align}
Then, if $c$ is positive, $Q[\partial_t]$ is negative,
and this suggests that $Q[\partial_t]$ is not physically reasonable as a definition of the mass.
On the other hand, we can show that the ADM momentum, defined in \eqref{eq:ADM-def},
is independent of the parameter, $c$,
\begin{align}
 P^{(\mathrm{string})}_A
 = n_{\scriptscriptstyle{D}}\,(0,-1,0\dotsc,0\,;\,+1,0,\dotsc,0) \,,
\end{align}
and the ADM energy, $P^{(\mathrm{string})}_t$, is always positive.

\end{itemize}

\subsection{Non-Riemannian Geometry T-dual to Fundamental String}
\label{sec:Non-Riemannian}

Now, in order to obtain a non-Riemannian  background,
we further perform  double $T$-duality transformations along { the isometric} $t$- and $z$-directions (see (5.23) in \cite{Lee:2013hma} for the explicit form).
Through the $\ODD$ rotations~\footnote{
Note that the $\ODD$ rotation here may not correspond to the traditional T-duality rotation. In backgrounds with isometries, we can choose the coordinates, $x^A=(\tilde{x}_a,x^a,\tilde{x}_i,x^i)$ ($a=1,\dotsc,D-n$, $i=D-n+1,\dotsc,D$), such that the background fields are independent of $\tilde{x}_a$ and $x^I=(\tilde{x}_i,x^i)$. In such backgrounds, a global $\ODD$ rotation, $\cH_{AB} \rightarrow O_A{}^C\,O_B{}^D\, \cH_{CD}$ with $O_A{}^B=\bigl(\begin{smallmatrix} 1 & 0 \cr 0 & O_I{}^J \end{smallmatrix}\bigr) \in \ODD$ (keeping the coordinates fixed), transforms the equation of motion of DFT covariantly.  We used this rotation as a solution generating method. For discussions of related subtle issues, see \cite{Cederwall:2014kxa,Cederwall:2014opa}.}, the $(t,z,\tilde{t},\tilde{z})$ part of
the generalized metric, $\cH_{AB}$, has the following form:
\begin{align}
 \left(\begin{smallmatrix}
 c\,(2+c\,H) & 0 & 0 & -(1+c\,H) \cr
 0 & -c\,(2+c\,H) & -(1+c\,H) & 0 \cr
 0 & -(1+c\,H) & -H & 0 \cr
 -(1+c\,H) & 0 & 0 & H \end{smallmatrix}\right)\,.
\end{align}
Then, in the $c\to 0$ limit, the upper-left $(2\times 2)$ block vanishes which would correspond to the inverse of the Riemannian metric. Namely, this background becomes singular in the conventional Riemannian  sense,
and accordingly is called a non-Riemannian background \cite{Lee:2013hma}.

In this background, from the asymptotic form
\begin{align}
 K^{tr}[\partial_t]\approx -c\,\partial_r H(r)\,,\quad
 K^{tr}[\tilde{\partial}^z]\approx -c^2\,\partial_r H(r)\,,\quad
 B^r \approx 0\,,
\end{align}
we obtain the ADM $2D$-momentum for the non-Riemannian (n-R) background,%
\footnote{For the non-Riemannian case, in order to calculate the ADM momentum explicitly,
we need to use a general formula \eqref{eq:Komar-munu-gen}, instead of \eqref{eq:Komar-munu}.
Further, since we cannot define $b_{\mu\nu}$ for the non-Riemannian case,
the ADM momentum should be defined using a different parametrization the generalized metric, see \textit{e.g.~}\cite{Sakatani:2014hba}.}
\begin{align}
 Q^{(\text{n-R})}[\partial_A]
 &= n_{\scriptscriptstyle{D}}\,(0,c^2,0\dotsc,0\,;\,c,0,0,\dotsc,0) \,,
\\
 P^{(\text{n-R})}_A
 &= n_{\scriptscriptstyle{D}}\,\Bigl(0,c^2,0\dotsc,0\,;\,\frac{c}{2+c},0,0,\dotsc,0\Bigr) \,.
\end{align}
In this case, the ADM energy, which depends on the parameter, $c$, can have a negative value.

\subsection{Reissner-Nordstr\"om Black Hole}
\label{sec:RN-BH}

As a simplest example in the YM-coupled DFT,
let us consider the Reissner-Nordstr\"om black hole,
\begin{align}
 \rmd s^2 &= -f(r)\, \rmd t^2 +f^{-1}(r)\, \rmd r^2 + r^2\, \rmd \Omega^2_{{\scriptscriptstyle{D{-2}}}}\,, \quad
 f(r) := 1-\frac{2\mu}{r^{D-3}} + \frac{q^2}{r^{2(D-3)}} \,, \quad
 \Phi =0\,,
\\
  A &= -\frac{2\gYM Q}{r^{D-3}}\,\rmd t\,, \quad B_{\mu\nu}=0\,,\quad
 \mu := \frac{M}{2(D-2)\,\Omega_{{\scriptscriptstyle{D{-2}}}}}\,,\quad
 q^2 := \frac{2(D-3)\,Q^2}{(D-2)} \,.
\end{align}
The doubled vector potential satisfies (\ref{secconV}) and  is parametrized by $V_{A}=(0,\,A_\mu)$.  Consequently, we have $(P\cF\brP)^{[\mu\nu]}=-f^{\mu\nu}=-G^{\mu\rho}G^{\nu\sigma}(\partial_{\rho}A_{\sigma}-\partial_{\sigma}A_{\rho})$, see (3.18) of \cite{Jeon:2011kp}.

In this background, the global charge~\eqref{MAINFINAL} becomes
\begin{align}
 Q[X]=2\,\int_{\partial\cM}\rmd^{{\scriptscriptstyle{D{-2}}}} x_{tr}\,\sqrt{|G|}\,
 \bigl(K^{[tr]}+2\,\xi^{[t}B^{r]} -4\,\gYM^{-2}\,f^{tr}A_t\,\xi^t\bigr) \,.
\end{align}
From the asymptotic behaviour, $f^{tr}A_t \propto r^{-(2D-5)}$,
the last term does not contribute to the surface integral.
The nontrivial contributions come from
\begin{align}
 K^{tr}[\partial_t]\approx 2 (D-3)\,\mu\,r^{-(D-2)} \,,\quad
 B^r \approx 2\,\mu\,r^{-(D-2)}\,,
\end{align}
such that we can recover the correct ADM mass,%
\footnote{If we make a constant shift in the gauge field, $A_\mu\to A_\mu + a_\mu$,
the $f^{tr}A_t$ term also gives a contribution to the ADM mass;
$Q[\partial_t]\to Q[\partial_t]-8(D-3)\,\Omega_{{\scriptscriptstyle{D{-2}}}}\,\gYM^{-1}\,Q\,a_t$, which depends on the free parameter, $a_t$. Like the definition of the ADM momentum given in \eqref{eq:ADM-def}, we can define a gauge invariant combination, $\widehat{P}_t := P_t +8(D-3)\,\Omega_{{\scriptscriptstyle{D{-2}}}}\,\gYM^{-1}\,Q\,A_t$.}
\begin{align}
 P_t = Q[\partial_t] = 2 (D-2)\,\Omega_{{\scriptscriptstyle{D{-2}}}}\,\mu = M \,.
\end{align}

\subsection{Black Five-Brane}
\label{sec:5-brane}

Here, we consider the black 5-brane, whose background reads \cite{Horowitz:1991cd}:
\begin{align}
 \rmd s^2 &= -\frac{\bigl(1-\frac{r_+^2}{r^2}\bigr)}{\bigl(1-\frac{r_-^2}{r^2}\bigr)}\,\rmd t^2
             + \frac{\rmd r^2}{\bigl(1-\frac{r_+^2}{r^2}\bigr)\,\bigl(1-\frac{r_-^2}{r^2}\bigr)}
 +r^2\,\rmd\Omega_3^2 + \sum_{s=1}^5 (\rmd z^s)^2 \,,
\\
 e^{-2\Phi} &= 1-\frac{r_-^2}{r^2}\,,\quad
 \rmd B = Q\,\boldsymbol{\epsilon}_3 \,,\quad Q:= r_+r_- \,,
\label{eq:5-brane}
\end{align}
where $\boldsymbol{\epsilon}_3$ is the volume element on the unit 3-sphere, satisfying
$\int_{S^3}\boldsymbol{\epsilon}_3=2\pi^2$, and
$r_\pm$ is the radius of the outer and the inner horizons.
In order to make the conserved charges finite,
we assumed that the $z^s$-directions to be a five-torus with the volume, $V_{T^5}$.
In the extremal limit, $r_+\to r_-$, this background approaches that of the NS5-brane.

In the Cartesian coordinates, we obtain the asymptotic form,
\begin{align}
 K^{tr}[\partial_t] \approx \frac{2\,(r_+^2-r_-^2)}{r^3}\,,\quad
 B^r \approx \frac{r_+^2 + r_-^2}{r^3} \,.
\end{align}
As it is expected, the only non-vanishing component of the ADM momentum is the ADM energy,
\begin{align}
 P_t = Q[\partial_t]
 = \int\rmd^5 z\int_{S^3_\infty}\rmd\Omega_3 \, \bigl(3r_+^2-r_-^2\bigr)
 =V_{T^5}\, 2\pi^2\, \bigl(3r_+^2-r_-^2\bigr) \,,
\end{align}
which reproduces the known result, (3.14) in \cite{Lu:1993vt},
especially the mass of the NS5-brane as the extremal limit, $r_\pm\to N^{1/2}l_s$.
Note that the ADM momentum is timelike in this background,
\begin{align}
 (\cH^{(0)})^{AB}\,P_A\,P_B < 0 \,.
\end{align}
As expected, unlike the case of the fundamental string,
the charge of the NS5-brane does not appear as the ADM momentum.
The charge of the NS5-brane will appear as the NUT charge
in doubled spacetime since it is $T$-dual to the Kaluza-Klein monopole, which has the NUT charge.
Another possibility is that, as discussed in \cite{Berman:2014jsa,Berman:2014hna},
since monopoles are simultaneously interpreted as null waves in the `Exceptional Field Theory',
it may be possible to describe the charge of the NS5-brane
as an ADM momentum in the extended spacetime.

\subsection{Linear Dilaton Background}
\label{sec:linear-dilaton}

Finally, we demonstrate that our general formula \eqref{defQ}
is also applicable for a non-asymptotically flat background by adding a suitable counterterm as a boundary action.

As an example of a non-asymptotically flat background,
consider the asymptotically linear dilaton background,
which can be obtained by taking a decoupling limit
and performing a coordinate transformation \cite{Maldacena:1997cg}:
\begin{align}
 \rmd s^2 &= -f(r)\,\rmd t^2 + \frac{N l_s^2}{r^2f(r)} \,\rmd r^2
           + N l_s^2\, \rmd\Omega_3^2
           + \sum_{s=1}^5 (\rmd z^s)^2 \,,
\\
 e^{2\Phi} &= \frac{N l_s^2}{r^2} \,, \quad
 \rmd B = Q\,\boldsymbol{\epsilon}_3 \,,\quad
 f(r):=1-\frac{r_0^2}{r^2} \,.
\end{align}
Here, $r_0$ is the corresponds to the outer horizon and
$N$ corresponds to the number of the five-branes.

In this case, we have
\begin{align}
 K^{tr}[\partial_t] = \frac{r^2}{Nl_s^2}\,\partial_r f(r)\,,\quad
 B^r =  -\frac{4r}{Nl_s^2}\,f(r) - \frac{r^2}{Nl_s^2}\,\partial_r f(r) \,.
\end{align}

Since the constant term in $B^r$ gives a divergent value to the conserved charge, we add the following boundary term to the action:
\begin{align}
 S_0 = -\int_{\mathbb{R}_{t}\times\partial\cM} \sqrt{h}\,e^{-2\Phi}\, b_0 \, . 
\end{align}
Here, $h$ is the induced metric on the boundary, $\mathbb{R}_{t}\times\partial\cM$,
which is in our case a constant $r$ surface,
and $b_0$ is a function of $h$.
This corresponds to the shift in $B^A$ by
\begin{align}
 B^A \to \bar{B}^A:=B^A + b_0\,\hat{n}^A \,,
\label{eq:B-shift}
\end{align}
where $\hat{n}^A:=\cH^{AB}\,n_A$  and $n_{A}$ denote the unit normal vector at the boundary,
$\cH^{AB}{n}_A{n}_B=1$.%
\footnote{Note that $n_A$ is defined through Stokes' theorem for an arbitrary vector, $K^{A}$,
\begin{align*}
 \int \rmd^{{\scriptscriptstyle{D}}}x\, \partial_A \bigl(e^{-2d}K^A\bigr)
 = \oint \rmd^{{\scriptscriptstyle{D{-1}}}}x\, e^{-2d} n_A\, K^A \,.
\end{align*}
We note that $n^A=\cJ^{AB}\,n_B$ is different from $\hat{n}^A$ appearing in \eqref{eq:B-shift}.
Similarly, the normal vector, $\hat{n}^k\partial_k$, appearing in \eqref{eq:ADM-mass}
should be also understood as the $D$-dimensional components of $\hat{n}^A$.}
In the present case, its asymptotic form becomes $\hat{n}^A\partial_A \approx \partial_r$.
Thus, the shift simply changes $B^r$ component as
\begin{align}
 B^r \to \bar{B}^r = B^r + b_0 \,.
\end{align}
Here, we simply choose $b_0$ as the minus of the leading term in $B^r$,
evaluated on the extremal background, $r_0=0$; $b_0= 4/(N^{1/2}l_s)$.
We then obtain
\begin{align}
 K^{tr}[\partial_t] + \bar{B}^r
 \approx \frac{2r_0^2}{Nl_s^2}\,\frac{1}{r} \,.
\end{align}
Therefore, the ADM mass, which is the only non-vanishing component of the ADM momentum, becomes
\begin{align}
 Q[\partial_t] = \int\rmd^5z \int_{S^3_\infty}\rmd \Omega_3\, e^{-2d}\bigl(K^{tr}[\partial_t] + \bar{B}^r\bigr)
 = 2r_0^2 \, \Omega_3\,V_{T^5} \,.
\end{align}
This result matches with (138) in \cite{Clement:2004ii},
where the mass was obtained from an approach of  Brown and York~\cite{Brown:1992br} as well as of Hawking and Horowitz~\cite{Hawking:1995fd}. See also (6.17) of  \cite{Mann:2009id}, where another approach was used.

\section{Discussion}
In this paper, we formulated the conserved Noether current and associated global charges of the massless sector of string theory in the DFT approach. The result is manifestly $\ODD$-covariant. We checked the result against various string theory backgrounds, not only geometric but also non-geometric, and found that the result yields the right answers in all cases.

There are further directions our result can be extended. One would like to include the R-R sector fields and also to substantiate fermionic Noether currents and fermionic global charges as odd-grading part of supersymmetric asymptotic symmetries in superstring theories.

The most interesting and important applications of our result would be for non-geometric backgrounds, either from exotic branes or exotic fluxes. As a step torward this goal, we also apply our formula to the exotic $5^2_2$-brane background \cite{LozanoTellechea:2000mc,deBoer:2010ud,deBoer:2012ma},
\begin{align}
 \rmd s^2 &= H(r)\,\bigl(\rmd r^2+r^2\,\rmd \theta^2\bigr)
 + H(r)\,K^{-1}(r,\theta)\,\rmd x_{89}^2
 + \rmd x_{03\cdots 7}^2 \,,\quad
 H(r):= \sigma\,\ln(r_c/r) \,,
\\
 e^{2\Phi}&= H(r)\,K^{-1}(r,\theta)\,,\quad B_{89}=- K^{-1}(r,\theta)\,\theta\,\sigma\,, \quad
 K(r,\theta):= H^2(r) + \sigma^2\,\theta^2 \,,
\end{align}
where $\sigma:=R_8R_9/(2\pi l_s^2)$\,.
In this background, we obtain
\begin{align}
 K^{tr} = 0\,,\quad B^r = - \partial_r H^{-1}(r) \,,
\end{align}
and the ADM mass is obtained as
\begin{align}
 Q[\partial_t] = -\int_{T_{3456789}} \rmd^7z \int\rmd\theta\, r\,H^{-1}(r)\,\partial_r H(r)
 = 2\pi\sigma\,V_{T_{3456789}}\, H^{-1}(r)\,.
\end{align}
This agrees with the result in \cite{deBoer:2010ud,deBoer:2012ma}, and can be viewed alternative derivation starting from our manifestly $\ODD$ covariant formulation. If we follow the ad hoc procedure in \cite{deBoer:2010ud,deBoer:2012ma}, $H(r)\to 1$ as $r\to\infty$, we obtain the known ADM mass for the $5^2_2$-brane background,
\begin{align}
 Q[\partial_t] = 2\pi\sigma\,V_{T_{3456789}}
 = l_s^{-1}\,\frac{R_3\cdots R_7(R_8R_9)^2}{g_s^2\, l_s^9} \,.
\end{align}

We intend to report our further progress in the above directions in future works. \\

\noindent\textbf{Acknowledgements}. We thank Dongsu Bak, Masafumi Fukuma, Euihun Joung, Wontae Kim and Sang-Heon Yi for helpful discussions.  We also thank anonymous referee for constructive suggestions. This work was  supported in part by the National Research Foundation of Korea through the Grants   2013R1A1A1A05005747 and 2015K1A3A1A21000302 for JHP, and through the Grants 2005-0093843, 2010-220-C00003 and 2012K2A1A9055280 for SJR, WR and YS.

\newpage
\appendix


\section{DFT-Noether potential}
\label{app:Noether}

Here, we obtain explicit expressions for the DFT-Noether potential defined in \eqref{Kpotential}:
\begin{align}
 K^{AB}[X] = 4\,\bigl(\brP^{C[A}\,P^{B]D} -P^{C[A}\,\brP^{B]D}\bigr)\,
 \bigl(\partial_C X_D + \Gamma_{CDE} \,X^E \bigr) \,.
\end{align}
Using the identity,
\begin{align}
 &4\,\bigl(\brP^{AC}\,P^{BD} -P^{AC}\,\brP^{BD}\bigr)\, \Gamma_{CDE}
\nonumber\\
 &=8\,\bigl(\brP^{AC}\,P^{BD} -P^{AC}\,\brP^{BD}\bigr)\,
 \bigl[ \bigl(P\,\partial_C P\brP\bigr)_{[DE]}
+\bigl({\brP}_{[D}{}^F\,{\brP}_{E]}{}^G - P_{[D}{}^F\, P_{E]}{}^G \bigr)\,\partial_F P_{GC} \bigr]
\nonumber\\
 &= 4 \bigl[2\,\brP^{C[A}\,\bigl(P\,\partial_C P\brP\bigr)^{B]}{}_E
            +2\,  P^{C[A}\,\bigl(P\,\partial_C P\brP\bigr)_E{}^{B]}
            -\cH_E{}^C\,\bigl(P\,\partial_C P\brP\bigr)^{[AB]}
            + \bigl(P\,\partial_E P\brP\bigr)^{(AB)} \bigr]
\nonumber\\
 &= - 2\,\cH^{C[A}\,\cH^{B]D}\,\partial_C\cH_{DE}
    + 2\,\partial^{[A}\cH^{B]}{}_E
    - \cH_E{}^C\,\cH^{[A}{}_D\,\partial_C\cH^{B]D}
    + \partial_E\cH^{AB} \,,
\end{align}
the DFT-Noether potential becomes
\begin{align}
 K^{AB}[X] &= -2\,\cH^{C[A}\,\bigl(\partial_C X^{B]} + \partial^{B]} X_C\bigr)
 - 2\,\cH^{C[A}\,\cH^{B]D}\,\partial_C\cH_{DE}\,X^E
\nonumber\\
 &\quad + 2\,\partial^{[A}\cH^{B]}{}_E\,X^E  - \cH_E{}^C\,\cH^{[A}{}_D\,\partial_C\cH^{B]D}\,X^E \,.
\end{align}
Then, further using a parametrization \eqref{eq:X-param}
and the strong constraint, $\tilde{\partial}^\mu =0$, we obtain
\begin{align}
 K^{\mu\nu}[X]
 = -2\, \cH^{\rho[\mu}\, \bigl(\partial_\rho \xi^{\nu]}
  +\cH^{\nu]D}\,\partial_\rho \cH_{DE}\,X^E \bigr)
 - \cH_E{}^\rho\,\cH^{[\mu}{}_D\,\partial_\rho\cH^{\nu]D}\,X^E \,.
\label{eq:Komar-munu-gen}
\end{align}
Lastly, from the parametrization of the generalized metric~\eqref{aPARAN},
we can obtain the expression of \eqref{eq:Komar-munu}.

\section{Fall-off behaviour at infinity in the asymptotically flat case\label{APPDEMON}}

Here, we show that the left-hand side of \eqref{Thetapr}, $e^{-2d}\widehat{\Theta}^{A}(d,P,\delta d,\delta P)$,
indeed vanishes in the asymptotically flat background.
Using the fall-off behaviour \eqref{eq:fall-off} and the explicit form of $e^{-2d}\widehat{\Theta}^{A}(d,P,\delta d,\delta P)$,
\begin{align}
 e^{-2d}\widehat{\Theta}^{A}(d,P,\delta d,\delta P)
 = - 4\,\partial_B(e^{-2d}\,\cH^{AB})\,\delta d + e^{-2d}\,\Gamma_{BC}{}^A\, \delta \cH^{BC} \,,
\end{align}
we obtain
\begin{align}
 e^{-2d}\,\widehat{\Theta}^{A}(d,P,\delta d,\delta P) = \mathcal{O}(r^{2\alpha-1}) \,.
\end{align}
Further, since the integral measure at the spatial infinity behaves as
$\rmd^{{\scriptscriptstyle{D{-2}}}}x_{AB}\sim \mathcal{O}(r^{D-2})$
for a background with coordinates \eqref{eq:BG1}
(which includes the Schwarzschild black hole),
assuming $\alpha=-(D-3)$, we obtain
\begin{align}
 2\,\int_{\partial\cM}\rmd^{{\scriptscriptstyle{D{-2}}}}x_{AB}\, e^{-2d}\, X^{[A}\, \widehat{\Theta}^{B]}(d,P,\delta d,\delta P)
 =\mathcal{O}(r^{-(D-3)}) \,.
\end{align}
On the other hand, for a background with coordinates \eqref{eq:BG2}
(which includes the $p$-brane or the null-wave background),
we have $\rmd^{{\scriptscriptstyle{D{-2}}}}x_{AB}\sim \mathcal{O}(r^{D-p-2})$.
Assuming $\alpha=-(D-p-3)$, we get
\begin{align}
 2\,\int_{\partial\cM}\rmd^{{\scriptscriptstyle{D{-2}}}}x_{AB}\, e^{-2d}\, X^{[A}\,\widehat{\Theta}^{B]}(d,P,\delta d,\delta P)
 =\mathcal{O}(r^{-(D-p-3)}) \,.
\end{align}
Therefore, in both asymptotically flat backgrounds,
$e^{-2d}\widehat{\Theta}^{A}(d,P,\delta d,\delta P)$
does not give any contribution to the global charge. This validates the approximation~(\ref{Approx}).

\section{Conserved global charge in Einstein frame}

Our formula \eqref{defQ} for the conserved global charge,
restricted on the section, can be summarized as
\begin{align}
 &Q[X]=\int_{\partial\cM}\rmd^{{\scriptscriptstyle{D{-2}}}}x_{\mu\nu} \sqrt{|G|}\,e^{-2\Phi}\,
 \bigl(K^{\mu\nu}[X]+2X^{[\mu}B^{\nu]}\bigr)\,,
\\
 &K^{\mu\nu}[X]= 2\, \xi^{[\mu;\nu]} - H^{\mu\nu\rho}\, \zeta_\rho \,,\quad
 B^\mu =2\,G^{\mu\nu}\,\bigl(2\partial_\nu \Phi -\partial_\nu \ln\sqrt{|G|}\bigr) - \partial_\nu G^{\mu\nu} \,,
\end{align}
where we parametrized the $\ODD$-covariant DFT field variables in terms of the conventional Riemannian fields in string frame.
In this appendix, we obtain the corresponding expression
in terms of the Einstein frame metric.

In order to obtain the expression, we use the definition of the Einstein frame metric,
\begin{align}
 G^{\text{(string)}}_{\mu\nu} = e^{\frac{4}{D-2}\Phi}G^{\text{(E)}}_{\mu\nu}\,,
\end{align}
to rewrite the DFT-Noether potential, $K^{\mu\nu}$, and $2X^{[\mu}B^{\nu]}$.
The DFT-Noether potential can be rewritten as
\begin{align}
 \left[2\, \xi^{[\mu;\nu]}-H^{\mu\nu\rho}\zeta_\rho\right]^{\text{(string)}}
 =e^{-\frac{4}{D-2}\Phi}\left[2\, \xi^{[\mu;\nu]}-e^{-\frac{8}{D-2}\Phi}H^{\mu\nu\rho}\zeta_\rho+\frac{8}{D-2}\xi^{[\mu}\partial^{\nu]}\Phi\right]^{\text{(E)}} \,,
\end{align}
where an extra dilaton term appeared inside the bracket.
On the other hand, rewriting of $2X^{[\mu}B^{\nu]}$ also produces an additional dilaton term,
and they are cancelled with each other.
We then obtain the expression of conserved global charge in the Einstein frame:
\begin{align}
 &Q[X]=\int_{\partial\cM}\rmd^{{\scriptscriptstyle{D{-2}}}}x_{\mu\nu} \sqrt{|G^{\text{(E)}}|}\,
 \bigl(K_{\text{(E)}}^{\mu\nu}[X]+2X^{[\mu}B_{\text{(E)}}^{\nu]}\bigr)\,,
\\
 &K_{\text{(E)}}^{\mu\nu}[X]:= \bigl[2\, \xi^{[\mu;\nu]} - e^{-\frac{8}{D-2}\Phi}\,H^{\mu\nu\rho}\, \zeta_\rho\bigr]^{\text{(E)}} \,,\quad
 B_{\text{(E)}}^\mu := - 2\,G_{\text{(E)}}^{\mu\nu}\, \partial_\nu \ln\sqrt{|G^{\text{(E)}}|} - \partial_\nu G_{\text{(E)}}^{\mu\nu} \,.
\end{align}


\end{document}